\documentclass[prc,aps,a4paper,groupedaddress,superscriptaddress,nofootinbib,showpacs
,preprintnumbers,onecolumn]{revtex4}
\usepackage{graphicx}
\usepackage{amsfonts}
\usepackage{amssymb}
\usepackage{natbib}
\usepackage{dcolumn}
\usepackage{bm}
\newcommand{\bwt}{\begin{widetext}}
\newcommand{\ewt}{\end{widetext}}
\newcommand{\beq}{\begin{equation}}
\newcommand{\eeq}{\end{equation}}
\newcommand{\bea}{\begin{eqnarray}}
\newcommand{\eea}{\end{eqnarray}}
\begin{document}
\title{Dense stellar matter with trapped neutrinos under strong magnetic fields}
\author{Aziz Rabhi}
\email{rabhi@teor.fis.uc.pt}
\affiliation{Centro de F\' {\i}sica Computacional, Department of Physics, University of Coimbra, 3004-516 Coimbra, Portugal} 
\affiliation{Laboratoire de Physique de la Mati\`ere Condens\'ee,
Facult\'e des Sciences de Tunis, Campus Universitaire, Le Belv\'ed\`ere-1060, Tunisia}
\author{Constan\c ca Provid\^encia}
\email{cp@teor.fis.uc.pt}
\affiliation{Centro de F\' {\i}sica Computacional, Department of Physics, University of Coimbra, 3004-516 Coimbra, Portugal} 
\date{\today}
\begin{abstract}
We investigate the effects of strong magnetic fields on the equation of state of dense stellar neutrino-free and neutrino-trapped matter. Relativistic nuclear models both with constant couplings (NLW) and with density dependent parameters (DDRH) and including hyperons are considered . It is shown that at low densities neutrinos are suppressed in the presence of the magnetic field. The magnetic field reduces the strangeness fraction of neutrino-free matter and increases the strangeness fraction of neutrino-trapped matter. The mass-radius relation of stars described by these equations of state are determined. The magnetic field makes the overall equation of state  stiffer and the stronger the field the larger the mass of maximum mass star and the smaller the baryon density at the center of the star.  As a consequence in the presence of strong magnetic fields the possibility that a protoneutron star evolves to a blackhole is smaller.
\end{abstract}
\maketitle

\section{Introduction}
Neutron stars with very strong magnetic fields of the order of
$10^{14}-10^{15}$ G are known as magnetars \cite{duncan,usov,pacz}. They are believed to be the sources of the intense $\gamma$ and X rays (for a review refer to \cite{harding}).  Although  until presently only  16 magnetars have been identified as short $\gamma$-ray repeaters or anomalous X-ray pulsars~\cite{sgr}, according to Ref.~\cite{kouve}, a fraction as high as 10\% of the neutron star population could be magnetars. Magnetars   are warm, young stars, $\sim$ 1 kyear old. 

The study of dense stellar matter under strong magnetic fields is, therefore of great 
interest. Our knowledge of neutron star composition 
and structure is still uncertain~\cite{glen00}. For densities below twice normal nuclear 
matter density ($\rho_0 \sim 0.153$ fm$^{-3}$), the matter consists only of
nucleons and leptons. For  baryon densities above $2 \rho_0$, 
the equation of state (EOS) and the composition of matter are much less certain
and the strangeness degree of freedom should be taken into account either
through the onset of hyperons, kaon condensation or a deconfinement phase transition into strange
quark matter.
The presence of hyperons in neutron stars, which has been studied by
many authors~\cite{sm96, phzs99, glen01, shen02, sz03, pmp04},
 tends to soften the EOS at high density and lower the 
maximum mass of neutron stars~\cite{sm96, phzs99,hh00, glen01, shen02, sz03,
  pmp04}. Cold dense stellar matter is neutrino free but the protoneutron
star formed after
a supernova explosion, with an entropy per baryon of the order of
1 to 2, contains trapped neutrinos. After 10 to 20 s, the star
stabilizes at practically zero temperature and no trapped neutrinos
are left \cite{prak97}.

In this article, we focus on the properties of hyperonic
matter, which is composed of a chemically equilibrated and
charge-neutral mixture of nucleons, hyperons, and leptons. 
 The meson-hyperon couplings
play an important role in the determination of the EOS and 
the composition of hyperonic matter. In the presence of strong magnetic
fields, the pressure and composition of matter can be affected
significantly~\cite{prak02}. In Ref.~\cite{prak02}, the authors have
investigated the effects of strong magnetic fields on the properties 
of neutron star matter including hyperons, but no neutrino trapping was considered. 
 It was found  that the EOS at high density could be significantly affected both by Landau
quantization and by the magnetic moment interactions, but only
for field strength $B > 5 \times 10^{18}$ G. It was also shown that  the threshold densities of hyperons can be significantly altered by strong magnetic field. 
Similar conclusions were obtained in \cite{shen08} where the strangeness was included through an antikaon condensation or in \cite{shen09} where not only hyperons but also the strange mesons $\sigma^*$ and $\phi$ were included in the EOS.

It is very interesting to
investigate the influence of strong magnetic fields on hyperonic
matter when neutrinos are trapped. Although, in reality we should consider
warm matter, it was shown \cite{cp69} that the effect of the inclusion of trapped
neutrinos is more important than the temperature effect. We expect,
therefore, that the main conclusions taken for $T=0$ will still be valid at
finite temperature.
For matter without a strong magnetic field,  it has been shown in Ref.~\cite{prak97,cp69}   that the EOS  with trapped
neutrinos  are harder than the  neutrino-free EOS, and,  the maximum
baryonic allowed mass of a stable star is higher when neutrinos are
trapped. As a consequence, after the neutrino outflow has occurred the most
massive stars may decay into a blackhole.

The effect of the magnetic field on the structure and composition of a neutron
star allowing quark-hadron phase transition with  trapped neutrinos has been
studied in Ref.~\cite{chakrabarty98}. 
They have concluded that the strong magnetic field makes the overall equation of
state softer giving rise to much smaller maximum mass configurations which
do not favor the formation of low mass blackholes.  Their conclusion about
the reduction of the maximum mass configuration is in contrast with the
result from other works \cite{prak02, mbap09} where an increase is predicted
due to the positive contribution of the magnetic field pressure to the total
EOS. In Ref.~\cite{chakrabarty98} the magnetic field pressure enters the total
EOS as a negative contribution, making the overall EOS softer.

This work is organized as follows: we make a brief review of the
formalism used for the hadron matter with and without trapped neutrinos. 
Next we present and discuss the results obtained for the equation of state 
of stellar matter and the  mass/radius properties of the corresponding 
compact star families, for several values of the magnetic field. 
At the end we will draw some conclusions.
 
\section{The formalism}
For the description of the EOS of neutron star matter, we employ a
field-theoretical approach. The Lagrangian density of the relativistic TW
model~\cite{fuchs,tw} is given by  
\bea
{\cal L}&=&\sum_{b}\bar{\Psi}_{b}\bigg[ i\gamma_{\mu}\partial^{\mu}-q_{b}\gamma_{\mu}A^{\mu}- 
m_{b}+\Gamma_{\sigma  b}\sigma \cr
&-&\Gamma_{\omega  b}\gamma_{\mu}\omega^{\mu}-\Gamma_{\rho  b}\tau_{3_{b}}\gamma_{\mu}\rho^{\mu}
-\frac{1}{2}\mu_{N}\kappa_{b}\sigma_{\mu \nu} F^{\mu \nu}\bigg ]\Psi_{b} \cr
&+&\sum_{l} \bar{\psi}_{l}\left(i\gamma_{\mu}\partial^{\mu}-q_{l}\gamma_{\mu}A^{\mu}
-m_{l}\right )\psi_{l} \cr
&+&\frac{1}{2}\partial_{\mu}\sigma \partial^{\mu}\sigma
-\frac{1}{2}m^{2}_{\sigma}\sigma^{2}
+\frac{1}{2}m^{2}_{\omega}\omega_{\mu}\omega^{\mu}
-\frac{1}{4}\Omega^{\mu \nu} \Omega_{\mu \nu}  \cr
&-&\frac{1}{4} F^{\mu \nu}F_{\mu \nu}
+\frac{1}{2}m^{2}_{\rho}\rho_{\mu}\rho^{\mu}-\frac{1}{4}  P^{\mu \nu}P_{\mu \nu}
\label{lan}
\eea
where $\Psi_{b}$ and $\psi_{l} $ are the baryon and lepton Dirac fields,
respectively. The index $b$ runs overs the eight lightest baryons $n$, $p$,
$\Lambda$, $\Sigma^-$, $\Sigma^0$, $\Sigma^+$,  $\Xi^-$ and $\Xi^0$
(neglecting the $\Omega^{-}$ and the $\Delta$ quartet, which appear only at
quite high densities, does not qualitatively affect our conclusions), and the sum on $l$ is over electrons and muons ($e^{-}$ and $\mu^{-}$).
$\sigma$, $\omega$, and $\rho$ represent the scalar, vector, and isovector-vector meson fields, which are exchanged for the description of nuclear interactions and $A^\mu=(0,0,Bx,0)$ refers to a external magnetic field along the z-axis. The baryon mass and isospin projection are denoted by $m_{b}$ and $\tau_{3_b}$, respectively. The mesonic and electromagnetic field strength tensors are given by their usual expressions: $\Omega_{\mu \nu}=\partial_{\mu}\omega_{\nu}-\partial_{\nu}\omega_{\mu}$, $P_{\mu \nu}=\partial_{\mu}\rho_{\nu}-\partial_{\nu}\rho_{\mu}$, and  
$F_{\mu \nu}=\partial_{\mu}A_{\nu}-\partial_{\nu}A_{\mu}$. The baryon
anomalous magnetic moments (AMM) are introduced via the coupling of the
baryons to the electromagnetic field tensor with $\sigma_{\mu
  \nu}=\frac{i}{2}\left[\gamma_{\mu}, \gamma_{\nu}\right] $ and strength
$\kappa_{b}=(\mu_b/\mu_N-q_bm_p/m_b)$. We neglect the AMM of the leptons in
this work, because their effect is very small as shown in Ref.~\cite{r9}. For the
electromagnetic field, only frozen-field configurations are considered 
and thus there is no
associated field equation.
The density dependent strong interaction couplings are denoted by
$\Gamma$, the electromagnetic couplings by $q$ and the baryons, mesons and
leptons masses by $m$. The parameters of the model are the nucleon mass
$M=939$ MeV, the masses of mesons $m_\sigma$, $m_\omega$, $m_\rho$ and the
density dependent coupling parameters which are adjusted in order to reproduce
some of the nuclear matter bulk properties and relations with the Dirac-Brueckner Hartree-Fock (DBHF) calculations~\cite{dbhf}, using the following parametrisation
\beq
\Gamma_{i}(\rho)=\Gamma_{i}(\rho_{0})f_{i}(x),\quad i=\sigma, \omega
\label{gam1}
\eeq
with
\beq
f_{i}(x)=a_{i}\frac{1+b_{i}\left(x+d_{i}\right)^{2}}{1+c_{i}\left(x+d_{i}\right)^{2}}
\label{gam2}
\eeq
where $x={\rho}/{\rho_{0}}$ and
\beq
\Gamma_{\rho}(\rho)=\Gamma_{\rho}(\rho_{0})\exp\left[ -a_{\rho}(x-1)\right] 
\eeq
with the values of the parameters $m_i$, $\Gamma_{i}$, $a_{i}$, $b_{i}$, $c_{i}$ and $d_{i}$, $i=\sigma, \omega, \rho$ given in \cite{tw}. The meson-hyperon couplings are assumed to be fixed fractions of the meson-nucleon couplings, $\Gamma_{i H}=x_{i H} \Gamma_{i N}$, where for each meson $i$, the values of $x_{i H}$ are assumed equal for all hyperons H. The values of $x_{i H}$ are chosen to reproduce the binding energy of the $\Lambda$ at nuclear saturation as suggested by Glendenning and Moszkwoski and are given in Table~\ref{table2}. 

For GM1 model, we add to the Lagrangian density, Eq.~(\ref{lan}), the scalar meson self-interaction terms 
$${\cal{L}}_{nl\sigma} =-\frac{1}{3}bm_n(g_{\sigma}\sigma)^3-\frac{1}{4}c(g_{\sigma}\sigma)^4,$$ where   $g_i=\Gamma_i$, $b$ and $c$ are two dimensionless parameters.

\begin{table*}
\caption{Static properties of the baryons considered in this study. The mass,
  electric charge  and strange charge of the baryon $b$ are denoted by $m_b$, $q_b$ and $q^b_s$, respectively. The baryonic magnetic moment is denoted by $\mu_b$ and the anomalous magnetic moment   by $\kappa_b=(\mu_b/\mu_N-q_bm_p/m_b)$, where $\mu_N$ is  the nuclear magneton.}
\label{table1}
\begin{ruledtabular}
\begin{tabular}{ c c c c c c}
baryon name & Mass & Charge & Strangeness & Magnetic moment & Anomalous magnetic moment \\
      b              &  (MeV) &  $q_{b}(e)$ & $q^s_b$ & $\mu_{b}/\mu_{N}$ &  $\kappa_{b}$  \\
\hline
       p           & 938.27 & 1 & 0 &  2.97 & 1.79 \\
       n           & 939.56 & 0 & 0 & -1.91 &-1.91 \\
\hline
$\Lambda^0$   & 1115.7 & 0 & -1& -0.61 &-0.61 \\
\hline
$\Sigma^+$ & 1189.4 &  1 & -1 &  2.46 &  1.67 \\
$\Sigma^0$  & 1192.6 &  0 & -1 &  1.61 &  1.61 \\
$\Sigma^-$   & 1197.4 & -1 & -1 & -1.16 & -038 \\
\hline
$\Xi^0$        & 1314.8  &  0 & -2 & -1.25 & -1.25\\
$\Xi^-$        & 1321.3  &  -1 & -2 & -0.65 & 0.06\\
\end{tabular}
\end{ruledtabular}
\end{table*}

\begin{table*}
\caption{Saturation properties of nuclear matter  and the  nucleon-meson coupling constants for the GM1 and TW models. The relative hyperon-meson coupling constants  used in the calculation are also given.}
\label{table2}
\begin{ruledtabular}
\begin{tabular}{ c c c c c c c c c c c c c}
 & $\rho_{0}$& -B/A & &$\Gamma_{\sigma N}/m_{\sigma}$ &$\Gamma_{\omega N}/m_{\omega}$&$\Gamma_{\rho N}/m_{\rho}$&  &  &  & &\\
Model  & (fm) &(MeV)& $M^{*}/M$ & (fm) & (fm) & (fm) &$x_{\sigma H}$&$x_{\omega H}$ & $x_{\rho H}$ &b &c \\
\hline
GM1 &0.153&16.30&0.70& 3.434 & 2.674 & 2.100 & 0.600 & 0.653 & 0.600 &0.002947 &-0.001070 \\
                                                 &     & && & &  & 0.800 & 0.913 & 0.800 & &\\
TW &0.153 &16.30 & 0.56 &3.84901& 3.34919 & 1.89354& 0.600 & 0.658 & 0.600 &0.0 &0.0\\
                                                   &      & &&&& & 0.800 & 0.905 & 0.800&& \\
\end{tabular}
\end{ruledtabular}
\end{table*}

The field equations of motion follow from the Euler-Lagrange equations. From the Lagrangian density in Eq.~(\ref{lan}), we obtain the following meson field equations in the mean-field approximation 
\bea
m^{2}_{\sigma} \sigma &=&\sum_{b}\Gamma_{\sigma b}\rho^{s}_{b}=\Gamma_{\sigma N}\sum_{b}x_{\sigma b}\rho^{s}_{b} \label{mes1} \\
m^{2}_{\omega} \omega^{0}   &=& \sum_{b}\Gamma_{\omega b}\rho^{v}_{b}=\Gamma_{\omega N}\sum_{b}x_{\omega b}\rho^{v}_{b} \label{mes2} \\
m^{2}_{\rho} \rho^{0} &=&\sum_{b}\Gamma_{\rho b}\tau_{3_{b}}\rho^{v}_{b}=\Gamma_{\rho N}\sum_{b}x_{\rho b}\tau_{3_{b}}\rho^{v}_{b} \label{mes3}
\eea
where $\sigma= \left\langle \sigma \right\rangle $, $\omega^{0}= \left\langle \omega^{0} \right\rangle $ and $ \rho^{0}= \left\langle \rho^{0} \right\rangle $ are the nonvanishing expectation values of the meson fields in uniform matter.

The Dirac equations for baryons and leptons are, respectively, given by 
\bea
\big[i\gamma_{\mu}\partial^{\mu}-q_{b}\gamma_{\mu}A^{\mu}-m^{*}_{b}
-\gamma_{0}\left(\Gamma_{\omega}\omega^{0}
+\Gamma_{\rho}\tau_{3_{b}}\rho^{0}+\Sigma^{R}_{0}\right) 
-\frac{1}{2}\mu_{N}\kappa_{b}\sigma_{\mu \nu} F^{\mu \nu}\big] \Psi_{b}&=&0 \label{MFbary}\\
\left(i\gamma_{\mu}\partial^{\mu}-q_{l}\gamma_{\mu}A^{\mu}-m_{l} \right) \psi_{l}&=&0 \label{MFlep}
\eea
where $ m^{*}_{b}=m_{b}-\Gamma_{\sigma}\sigma $ is the effective mass of baryon species $b$. 
For neutrino-free stellar matter consisting of a $\beta$-equilibrium  mixture of baryons and
leptons, the following equilibrium conditions must be imposed:
\bea \mu_b=q_b \, \mu_n - q_l \, \mu_e,\label{free}\eea
which are equivalent to
\bea
\mu_{n} &= &\mu_{\Lambda}=\mu_{\Sigma^0}=\mu_{\Xi^0}, \cr 
\mu_{p}&=&\mu_{\Sigma^+}=\mu_{n}-\mu_{e}, \cr
\mu_{\Sigma^-} &=& \mu_{\Xi^-}=\mu_{n}+\mu_{e}, \cr 
\quad\mu_{\mu}&=&\mu_{e},
\label{beta}
\eea
where $\mu_i$ is the chemical potential of species $i$. The electric charge neutrality condition is expressed by
\beq
\sum_{b} q_{b} \rho^{v}_{b}+\sum_{l}q_{l} \rho^{v}_{l}=0,
\label{neutra}
\eeq
where $ \rho^{v}_{i}$ is the number density of particle $i$.
If trapped neutrinos are included, we replace $\mu_{e}\rightarrow
\mu_{e}-\mu_{\nu_e}$ in the above equations,
\bea \mu_b=q_b \, \mu_n - q_l \, \left(\mu_e-\mu_{\nu_e}\right).\label{trap}\eea
For leptons we have
\beq
\mu_{\mu}-\mu_{\nu_\mu}=\mu_{e}-\mu_{\nu_e}.
\eeq
In the above equations $\mu_{\mu},\, \mu_{\nu_e}$ are the chemical potential,
respectively,  of the  muon and electron neutrinos.
The introduction of additional variables, the neutrino chemical potentials,
requires additional constraints, which we supply by fixing the lepton fraction
$Y_{Le}=Y_{e}+Y_{\nu_{e}}=0.4$ \cite{burrows86, prak97}.  Also, because no
muons are  present before and during the supernova explosion,  the constraint
$Y_{L\mu}=Y_{\mu}+Y_{\nu_{\mu}}=0$ must be imposed.

The energy spectra for charged baryons, neutral baryons  and leptons (electrons and muons) are given by
\bea
E^{b}_{\nu, s}&=& \sqrt{k^{2}_{z}+\left(\sqrt{m^{* 2}_{b}+2\nu |q_{b}|B}-s\mu_{N}\kappa_{b}B \right) 
^{2}}+\Gamma_{\omega b} \omega^{0}+\tau_{3_{b}}\Gamma_{\rho b}\rho^{0}+\Sigma^{R}_{0} \label{enspc1}\\
E^{b}_{s}&=& \sqrt{k^{2}_{z}+\left(\sqrt{m^{* 2}_{b}+k^{2}_{x}+k^{2}_{y}}-s\mu_{N}\kappa_{b}B 
\right)^{2}}+\Gamma_{\omega b} \omega^{0}+\tau_{3_{b}}\Gamma_{\rho b}\rho^{0}+\Sigma^{R}_{0}\label{enspc2} \\
E^{l}_{\nu, s}&=& \sqrt{k^{2}_{z}+m_{l}^{2}+2\nu |q_{l}| B}\label{enspc3}
\eea
where $\nu=n+\frac{1}{2}-sgn(q)\frac{s}{2}=0, 1, 2, \ldots$ enumerates the Landau levels (LL) of the fermions with electric charge $q$, the quantum number $s$ is $+1$ for spin up and $-1$ for spin down cases, and the rearrangement term is given by 
\bea
\Sigma^{R}_{0}&=&\sum_{b}\left(\frac{\partial \Gamma_{\omega b}}{\partial \rho}\rho^{v}_b\omega_{0}+\frac{\partial \Gamma_{\rho b}}{\partial 
\rho}\tau_{3_{b}}\rho^{v}_{b}\rho_{0}-\frac{\partial \Gamma_{\sigma b}}{\partial \rho}\rho^s_{b}\sigma\right) \cr
&=&\frac{1}{\Gamma_{\omega N}}\frac{\partial \Gamma_{\omega N}}{\partial \rho}m^{2}_{\omega}\omega^{2}_{0}+\frac{1}{\Gamma_{\rho N}}\frac{\partial \Gamma_{\rho N}}{\partial 
\rho}m^{2}_{\rho} \rho^2_{0}-\frac{1}{\Gamma_{\sigma N}}\frac{\partial \Gamma_{\sigma N}}{\partial \rho}m^{2}_{\sigma}\sigma^2.
\eea
For the charged baryons, we introduce the effective mass under the effect of a
magnetic field 
\beq
\bar m^c_{b}=\sqrt{m^{* 2}_{b}+2\nu |q_{b}|B}-s\mu_{N}\kappa_{b}B,
\label{mc}
\eeq
the expressions of the scalar and vector densities are, respectively, given by~\cite{broderick}
\bea
\rho^{s}_{b}&=&\frac{|q_{b}|Bm^{*}_{b}}{2\pi^{2}}\sum_{\nu=0}^{\nu_{\mbox{\small max}}}\sum_{s}\frac{\bar m^c_{b}}{\sqrt{m^{* 2}_{b}+2\nu |q_{b}|B}}\ln\left|\frac{k^{b}_{F,\nu,s}+E^{b}_{F}}
{\bar  m^c_{b}} \right|, \cr
\rho^{v}_{b}&=&\frac{|q_{b}|B}{2\pi^{2}}\sum_{\nu=0}^{\nu_{\mbox{\small max}}}\sum_{s}k^{b}_{F,\nu,s}. 
\eea
where $k^{b}_{F, \nu, s}$ is the Fermi momenta of charged baryons $b$ with quantum numbers $\nu$ and $s$. The Fermi energies $E^{b}_{F}$ are related to the Fermi momenta $k^{b}_{F, \nu, s}$ by
\beq
\left(k^{b}_{F,\nu,s}\right)^{2}=\left(E^{b}_{F}\right)^{2} -\left(\bar{m}^{c}_{b}\right)^{2}.
\eeq
For neutral baryons the Fermi momenta is denoted by $ k^{b}_{F, s}$, and the Fermi energy $E^ {b}_{F}$ is given by
\beq
\left(k^{b}_{F,s}\right)^{2} =\left(E^{b}_{F}\right)^{2} -\bar{m}^{2}_{b},
\eeq
with
\beq
\bar{m}_{b}=m^{*}_{b}-s\mu_{N}\kappa_{b}B.
\eeq
The scalar and vector densities of the neutral baryon $b$ are, respectively, given by
\bea
\rho^{s}_{b}&=&\frac{m^{*}_{b}}{4\pi^{2}}\sum_{s} \left[E^ {b}_{F}k^{b}_{F, s}-\bar{m}^{2}_{b}\ln\left|
\frac{k^{b}_{F,s}+E^{b}_{F}}{\bar{m}_{b}} \right|\right],  \cr
\rho^{v}_{b}&=&\frac{1}{2\pi^{2}}\sum_{s}\left[ \frac{1}{3}\left(k^{b}_{F, s}\right) ^{3}-\frac{1}
{2}s\mu_{N}\kappa_{b}B\left(\bar{m}_{b}k^{b}_{F,s}+\left(E^{b}_{F}\right)^{2}\left(\arcsin\left( \frac{\bar{m}_{b}}
{E^{b}_{F}}\right) -\frac{\pi}{2} \right)  \right) \right]. 
\eea
The vector densities for leptons are given by
\beq
\rho^{v}_{l}=\frac{|q_{l}|B}{2\pi^{2}}\sum_{\nu=0}^{\nu_{\hbox{\small max}}}\sum_{s}k^{l}_{F,\nu,s},
\eeq
where $k^{l}_{F, \nu, s}$ is the lepton Fermi momenta, which are related to the Fermi energy $E^{l}_{F}$ by
\beq
\left(k^{l}_{F,\nu,s}\right)^{2}=\left(E^{l}_{F}\right)^{2}-\bar{m}^{2}_{l}, \quad l=e, \mu, 
\eeq
with $\bar{m}^{2}_{l}=m^{2}_{l}+2\nu |q_{l}| B$.
The summation in $\nu$ in the above expressions terminates at $\nu_{max}$, the largest value of $\nu$ for which the square of Fermi momenta of the particle is still positive and which corresponds to the closest integer from below.
For neutrino-trapped matter, the neutrino density is given by 
\beq
\rho^{v}_{\nu_e} = \frac{{k^{{\nu_e}}_{F}}^3}{6 \pi^2}.
\eeq

The chemical potentials of baryons and leptons are defined as 
\bea
\mu_{b}&=& E^{b}_{F}+\Gamma_{\omega b}\omega^{0}+\Gamma_{\rho b}\tau_{3_{b}}\rho^{0}+\Sigma^{R}_{0} \\
\mu_{l} &=& E^{l}_{F}=\sqrt{\left(k^{l}_{F,\nu,s}\right)^{2}+\bar{m}^{2}_{l}}.
\eea

We solve the coupled Eqs.~(\ref{mes1})-(\ref{neutra}) self-consistently at a given baryon density $\rho=\sum_{b}\rho^{v}_{b}$ in the presence of strong magnetic fields. The energy density of neutron star matter is given by
\beq
\varepsilon_{m}=\sum_{b} \varepsilon_{b}+\sum_{l=e,\mu}\varepsilon_{l}+\frac{1}
{2}m^{2}_{\sigma}\sigma^{2}+\frac{1}{2}m^{2}_{\omega}\omega^{2}_{0}+\frac{1}{2}m^{2}_{\rho}\rho^{2}_{0},
\eeq 
where the energy densities of charged baryons and leptons have the following forms
\bea
\varepsilon_{b}&=&\frac{|q_{b}|B}{4\pi^ {2}}\sum_{\nu=0}^{\nu_{\mbox{\small max}}}\sum_{s}\left[k^{b}_{F,\nu,s}E^{b}_{F}+\left(\bar{m}^{c}_{b}\right) ^{2} 
\ln\left|\frac{k^{b}_{F,\nu,s}+E^{b}_{F}}{\bar{m}^{c}_{b}} \right|\right] ,\cr
\varepsilon_{l}&=&\frac{|q_{l}|B}{4\pi^ {2}}\sum_{\nu=0}^{\nu_{\mbox{\small max}}}\sum_{s}\left[k^{l}_{F,\nu,s}E^{l}_{F}
+\left(m^{2}_{l}+2\nu |q_{l}|B\right) 
\ln\left|\frac{k^{l}_{F,\nu,s}+E^{l}_{F}}{\sqrt{m^{2}_{l}+2\nu |q_{l}| B}} \right|\right],
\eea
while those of uncharged baryons are given by
\bea
\varepsilon_{b}&=&\frac{1}{4\pi^ {2}}\sum_{s}\bigg[\frac{1}{2}k^{b}_{F, s}\left(E^{b}_{F}\right)^{3} -\frac{2}
{3}s\mu_{N}\kappa_{b} B \left(E^{b}_{F}\right)^{3}\left(\arcsin\left(\frac{\bar{m}_{b}}{E^{b}_{F}} \right)-\frac{\pi}
{2}\right)-\left(\frac{1}{3}s\mu_{N}\kappa_{b} B +\frac{1}{4}\bar{m}_{b}\right) \cr
&&\left(\bar{m}_{b}k^{b}_{F, s}E^{b}_{F}+\bar{m}^{3}_{b}\ln\left|\frac{k^{b}_{F,s}+E^{b}_{F}}{\bar{m}_{b}} 
\right|\right) \bigg].
\eea
The pressure of neutron star matter can be obtained by 
\beq
P_{m}=\sum_{i}\mu_{i}\rho^{v}_{i}-\varepsilon_{m}=\mu_{n}\sum_{b}\rho^{v}_{b}-\varepsilon_{m}
\label{press}
\eeq
where the charge neutrality and $\beta$-equilibrium conditions are used to get the last equality.
If the stellar matter contains neutrinos trapped, their energy density and pressure contributions, respectively,
\bea
\varepsilon_{\nu_e}&=&\frac{{k^{\nu_e}_{F}}^4}{8 \pi^2}, \cr
P_{\nu_e}&=&\frac{{k^{\nu_e}_{F}}^4}{24 \pi^2},
\eea
should be added to the stellar matter energy and pressure.
The total energy density and pressure of the system includes the contribution of the magnetic field,
\bea
\varepsilon &=&\varepsilon_m+\frac{B^{2}}{2}, \cr
P &=& P_m+\frac{B^{2}}{2}.
\label{press1}
\eea

With the obtained EOS, the mass-radius relation and other relevant quantities of neutron star can be derived by solving the Tolman-Oppenheimer-Volkoff (TOV) equations.

\section{Results and discussion}

\subsection{Uniform magnetic field}
In this section we consider that the external magnetic field is constant. The magnetic field will be defined
in units of the critical field $B^c_e=4.414 \times 10^{13}$~G, so that $B=B^* \, B^c_e$. 
In order to study the effect of strong magnetic fields on the structure of hyperonic 
matter we use two different relativistic mean-field approaches: the GM1 parametrisation 
of the NLW models~\cite{gm91}, and the TW parametrisation of the density-dependent 
relativistic hadronic (DDRH) models~\cite{fuchs,tw}, given in Table~\ref{table2}. We include the baryonic octet in the EOS and choose two sets of hyperon-meson coupling constants given in Table~\ref{table2}. The static properties of the baryons considered are listed  in Table~\ref{table1}.

\begin{figure}[ht]
\vspace{1.5cm}
\centering
\includegraphics[width=0.85\linewidth,angle=0]{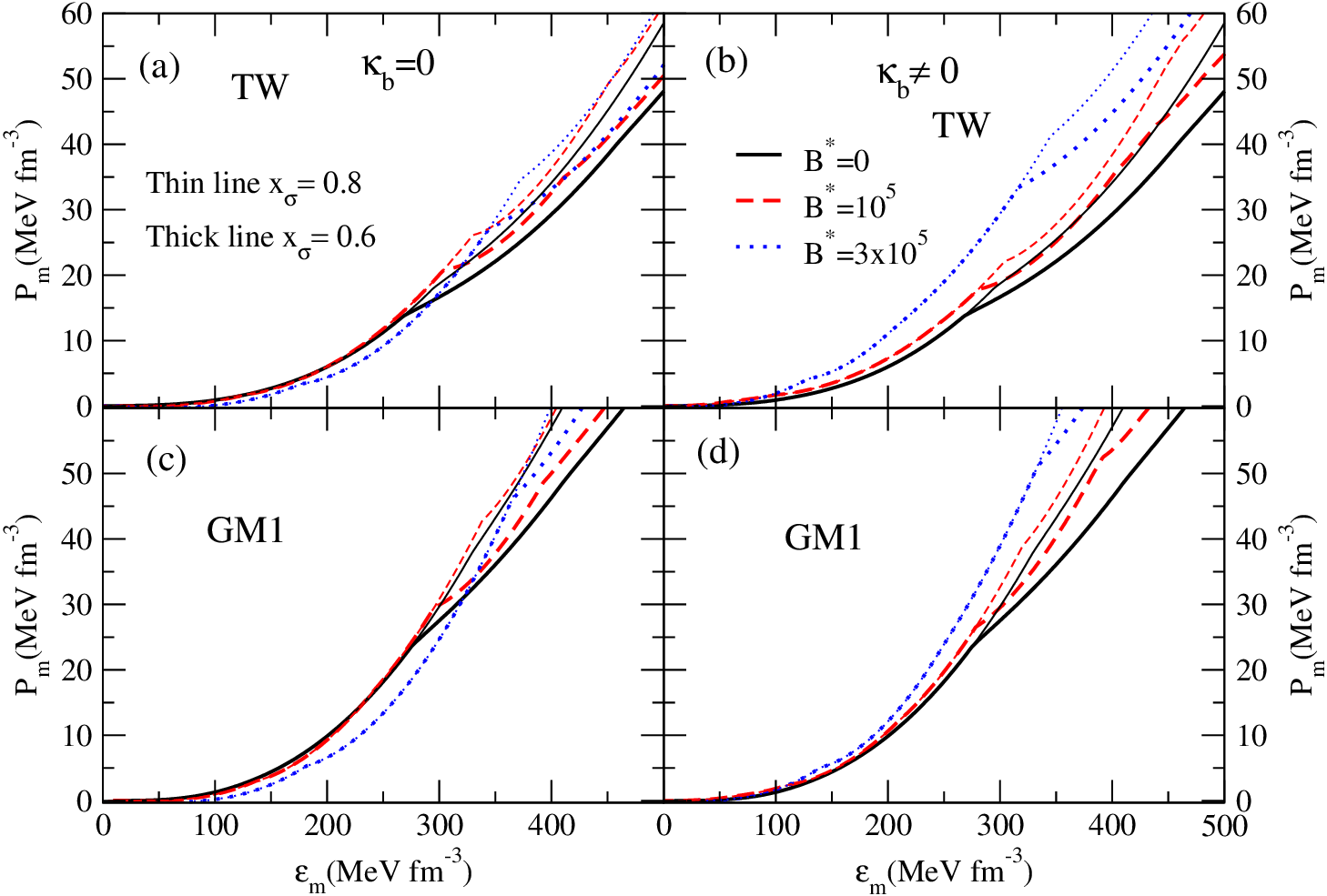}
\caption{EOS obtained with TW and GM1 models for neutrino free matter, and for several values of the magnetic field without (a) and (c) and with (b) and (d) AMM. Thick lines correspond to $x_\sigma=0.6$ and thin lines to $x_\sigma=0.8$. The kink in each EOS represents the onset of hyperons.}
\label{eosbc}
\end{figure}

In Fig.~\ref{eosbc},  the EoS obtained with both relativistic mean-field models GM1 
and TW are displayed for $B^*=0, 10^5, \:\hbox{and}\: 3\times10^5$, for the relative hyperon coupling constant $x_{\sigma}=0.6$ (thick lines) and $x_{\sigma}=0.8$ (thin lines). In Fig.~\ref{eosbc} (a) and Fig.~\ref{eosbc} (c) the EOS does not include the contribution form the AMM and in  Fig.~\ref{eosbc} (b) and Fig.~\ref{eosbc} (d) the AMM was included.  The kink on each one of the curves identifies the onset of hyperons.
Both for TW, which gives a softer EOS, and for GM1, the EOS at higher densities are harder for $x_{\sigma}=0.8$ than for $x_{\sigma}=0.6$ because a larger $\omega$ or $\rho$ coupling constant makes the onset of hyperons only possible at larger densities. This is also valid in the presence of a strong magnetic field.
The effects of the AMM  are only noticeable for the stronger magnetic field. It is seen that, in the absence of the AMM and in presence of a strong magnetic field, the EOS becomes softer for the smaller densities and harder at the larger densities due to the Landau quantization which affects charged particles.

\begin{figure}[ht]
\vspace{1.5cm}
\centering
\includegraphics[width=0.85\linewidth,angle=0]{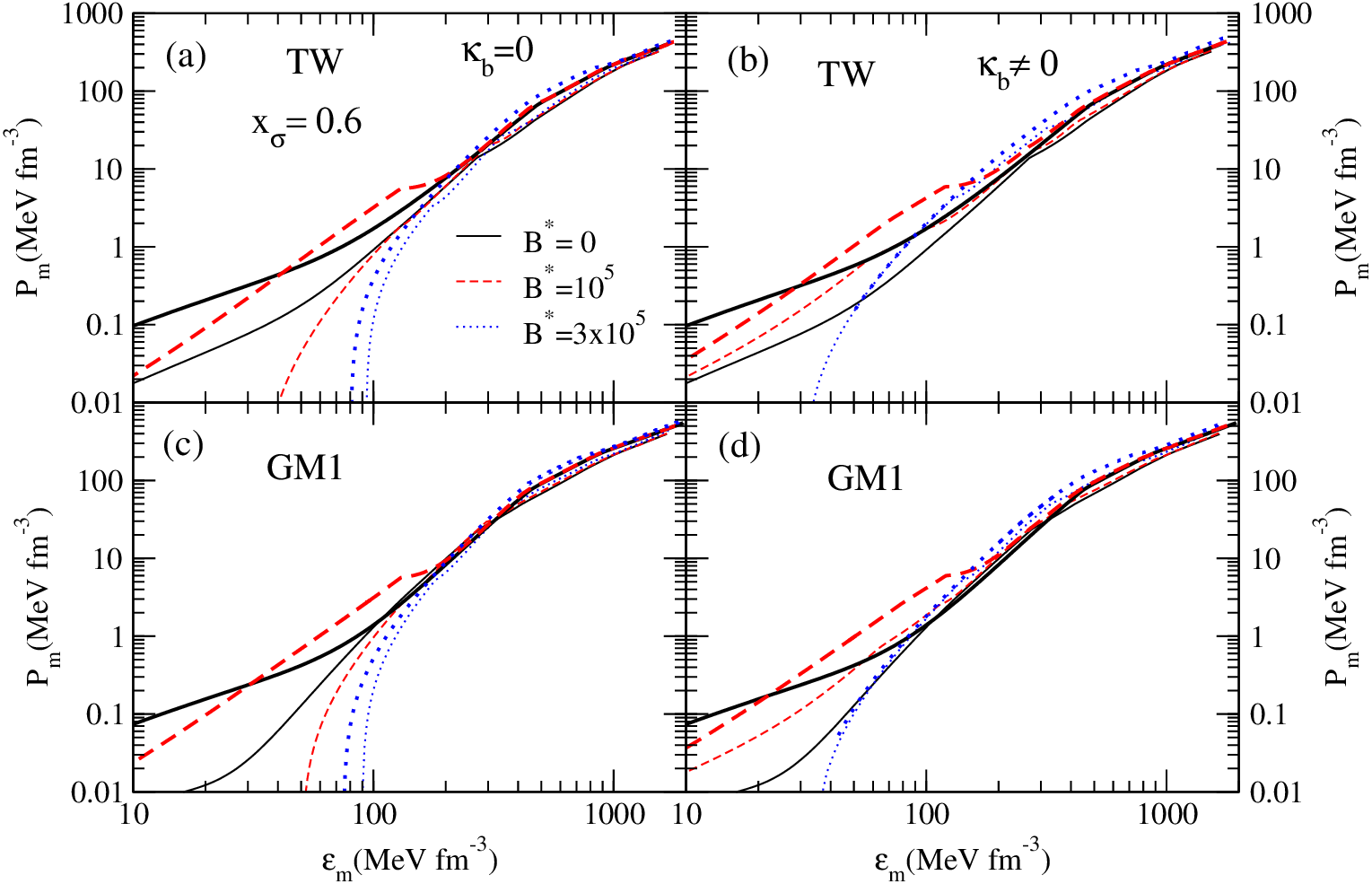}
\caption{EOS for TW and GM1 models and for $x_\sigma=0.6$, without (a) and (c) and with (b) and (d)  AMM. Thick lines for neutrino-trapped matter and thin lines for neutrino-free matter.}
\label{eosntvsnf}
\end{figure}

\begin{figure}[ht]
\vspace{1.5cm}
\centering
\includegraphics[width=0.85\linewidth,angle=0]{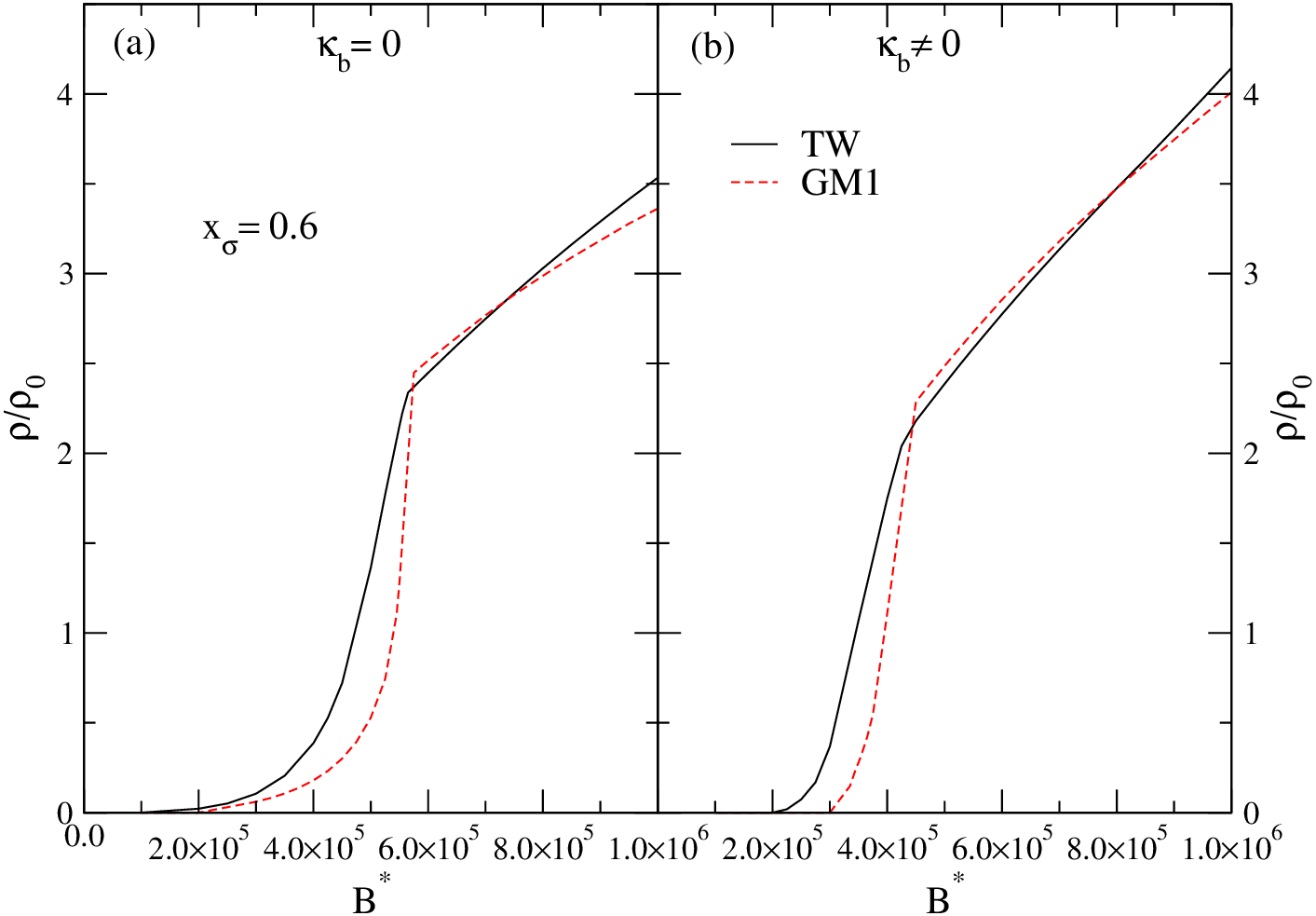}
\caption{Baryon density for which $\mu_{\nu_e}=0$ versus $B^*$ obtained for TW and GM1 models, with/without AMM and for $x_\sigma=0.6$. Below this density it is not possible to impose $Y_{Le}=0.4$.}
\label{Bvsdens}
\end{figure}

The effect of magnetic field on the EOS, when trapped neutrinos are considered, is seen in Fig.~\ref{eosntvsnf}  for GM1 and TW models, with $x_{\sigma}=0.6$ and without/with AMM. For $B=0$ the presence of neutrinos makes the EOS  harder~\cite{prak97, cp69}. A strong magnetic field without including the AMM, makes the EOS  harder at high densities if neutrino trapping is enforced. Moreover, if the AMM are included  the EOS become even harder.
At low densities, however,  the magnetic field softens the EOS even when the AMM are taken into account. We remark that for the largest field considered and taking into account AMM, it is not possible to get an EOS at low densities  for the lepton fraction $Y_L=0.4$.
In Fig.~\ref{Bvsdens} we plot the baryon density threshold, at which the neutrino chemical potential vanishes, as function of the magnetic field, for GM1 and TW models using $x_{\sigma}=0.6$ and without/with including AMM. Below this threshold density, it is not possible to impose  the lepton fraction $Y_L=0.4$.  The threshold density increases very fast with the increase of the magnetic field, \textit{e.g.} in Fig.~\ref{Bvsdens}(a) for $B^*=5\times10^5$ corresponds a threshold density equal to $0.5\rho_0$ for GM1 and $1.35\rho_0$ for TW. The kink at $\rho \sim 2.5\rho_0$ corresponds to the hyperon onset. However, at the surface the measured magnetic fields of magnetars are at most $B^* \sim 10^2$ and it is probable that even if the magnetic field is stronger in the interior it will be weaker than the threshold values given in Fig. \ref{Bvsdens}.
 
\begin{figure}[ht]
\vspace{1.5cm}
\centering
\includegraphics[width=0.85\linewidth,angle=0]{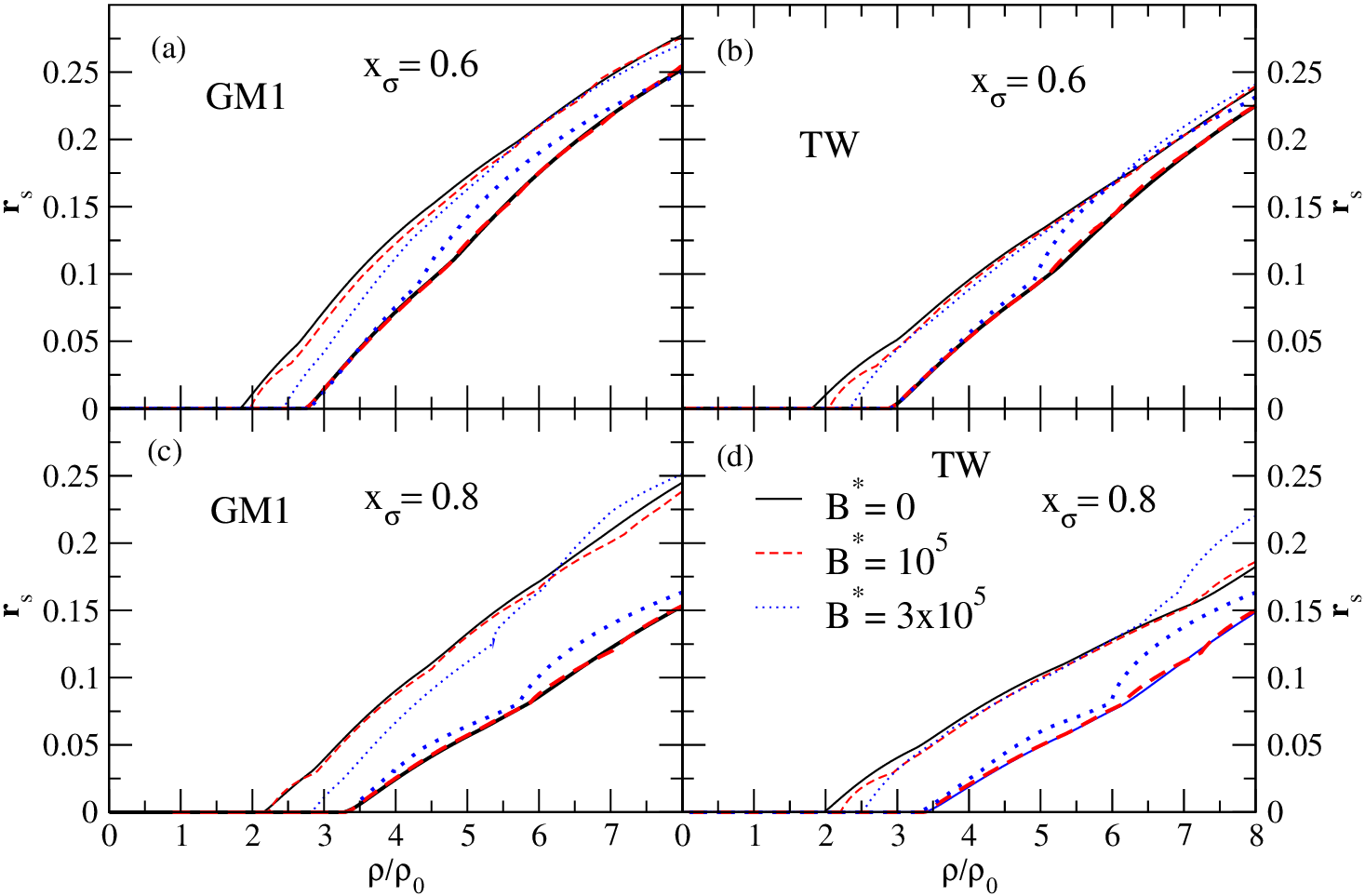}
\caption{Strangeness fraction as a function of the baryonic density, for severals values of B, without AMM, and for GM1 and TW models. (a) and (b) for $x_{\sigma}=0.6$ and (c) and (d) for $x_{\sigma}=0.8$. Thick lines correspond to matter with trapped neutrinos and thin lines to  neutrino-free matter.}
\label{strange1}
\end{figure}

In Fig.~\ref{strange1} we show, for several values of magnetic field without including the AMM, for GM1 and TW models and for the two values of the hyperon-meson coupling constants, the strangeness fraction defined as 
$$\displaystyle\mathfrak{r}^{HP}_s=\frac{\sum_{b}\left|q^{b}_ {s}\right|\rho_b}{3\rho},$$
 where $q^{b}_ {s}$ is the strange charge of baryon $b$, and is listed in Table~\ref{table1}.  
The strangeness onset occurs  around $2\rho_0$ and it has almost reached $0.275$ ($0.24$)  for GM1 and $0.23$ ($0.18$)  for TW, neutrino-free matter and  $x_{\sigma}=0.6$  ($x_{\sigma}=0.8$).  For $B=0$ and for the neutrino-free matter, increasing the hyperon-meson couplings decreases the strangeness contents, and this suppression of strangeness is stronger in TW than GM1. However, if the neutrino trapping is imposed the strangeness fraction is smaller and a stronger reduction occurs for GM1 and for $x_{\sigma}=0.8$.

For neutrino-free matter, the magnetic field reduces the strangeness fractions
below a critical density that can be quite high,  $6\rho_0$ or larger for the
fields shown. This effect starts to be detected already for $B^*=10^5$, which
represents the threshold for the magnetic field effects to become significant,
independently of the EOS~\cite{prak02}. However,
 the magnetic field enhances the strangeness fraction of neutrino-trapped matter. 
If the AMM are included the strangeness fraction behaves in the same way.
\begin{figure}[htb]
\vspace{1.5cm}
\centering
\includegraphics[width=0.85\linewidth,angle=0]{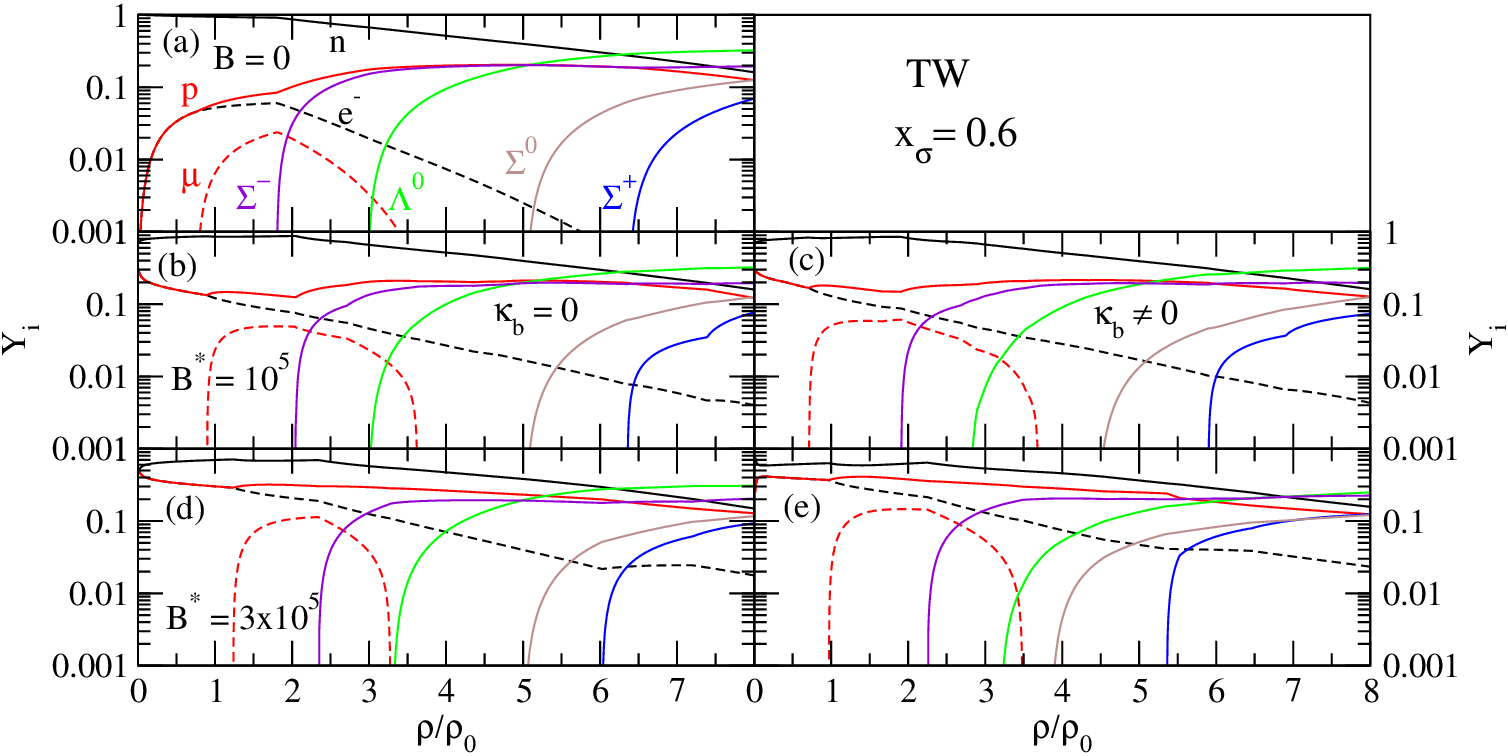}
\caption{Baryon fraction versus baryon density obtained with TW model for $x_{\sigma} = 0.6$ and neutrino-free matter.}
\label{frac06nf}
\end{figure}

\begin{figure}[htb]
\vspace{1.5cm}
\centering
\includegraphics[width=0.85\linewidth,angle=0]{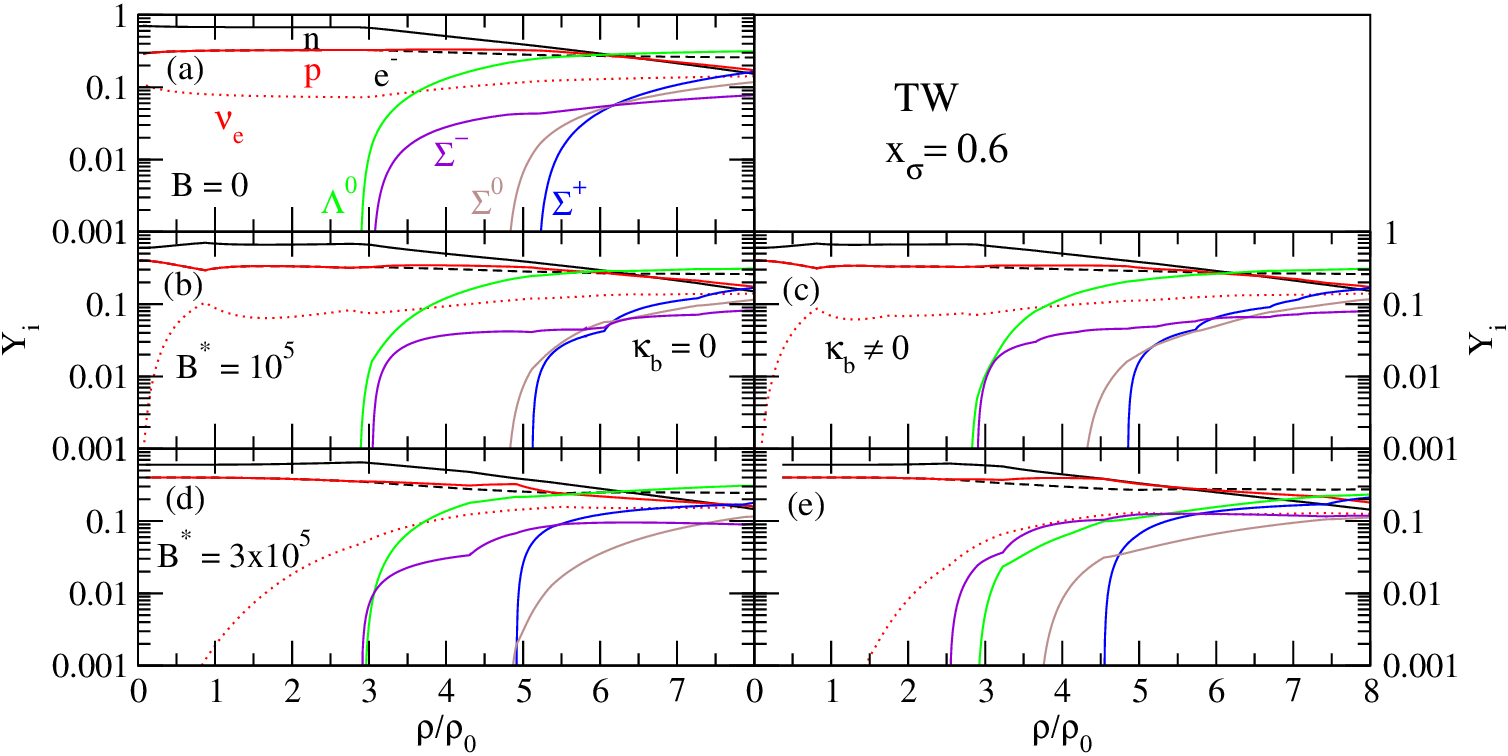}
\caption{Baryon fraction versus baryon density obtained with TW model for $x_{\sigma} = 0.6$, and neutrino-trapped matter.}
\label{frac06nt}
\end{figure}

\begin{figure}[htbb]
\vspace{1.5cm}
\centering
\includegraphics[width=0.85\linewidth,angle=0]{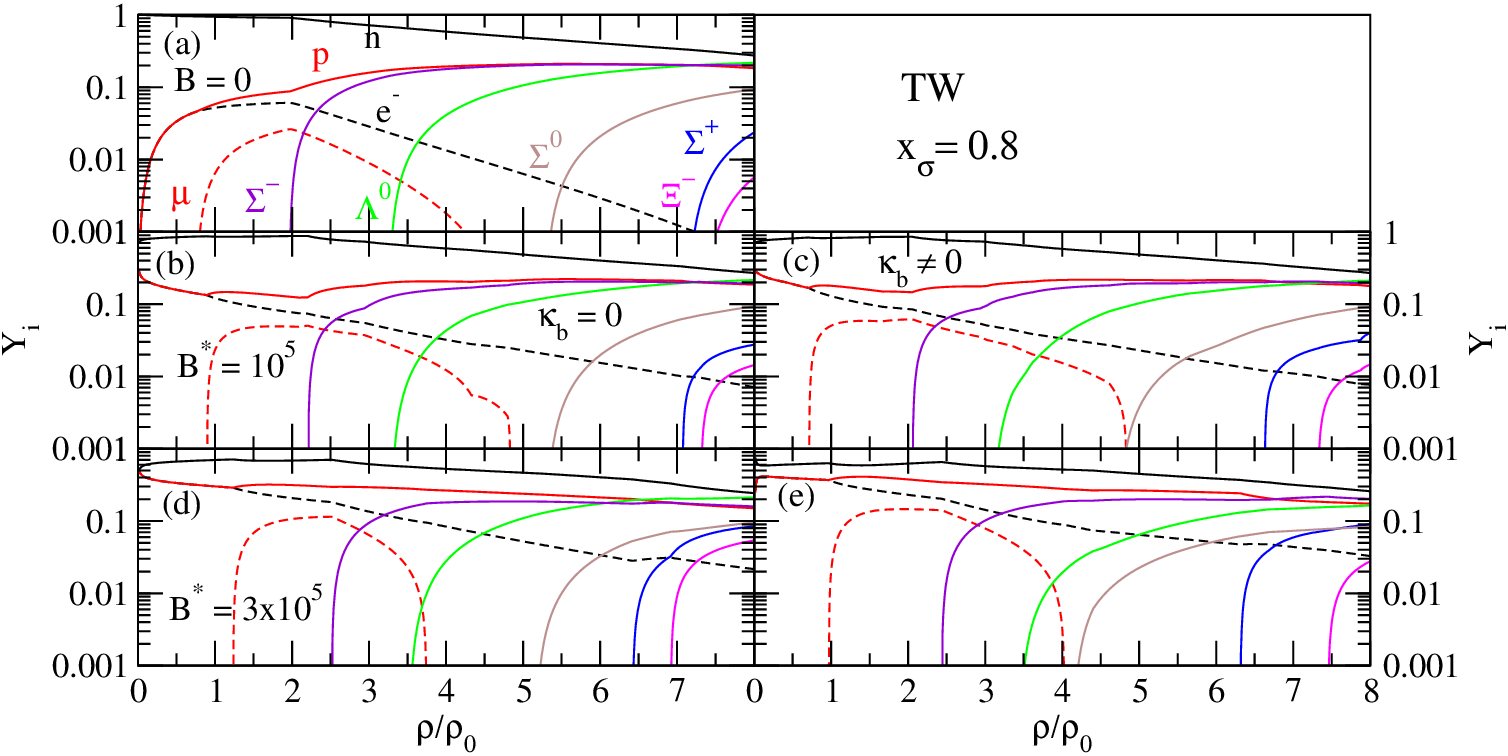}
\caption{Baryon fraction versus baryon density obtained with TW model, for $x_{\sigma} = 0.8$ and neutrino-free matter.}
\label{frac08nf}
\end{figure}

\begin{figure}[htb]
\vspace{1.5cm}
\centering
\includegraphics[width=0.85\linewidth,angle=0]{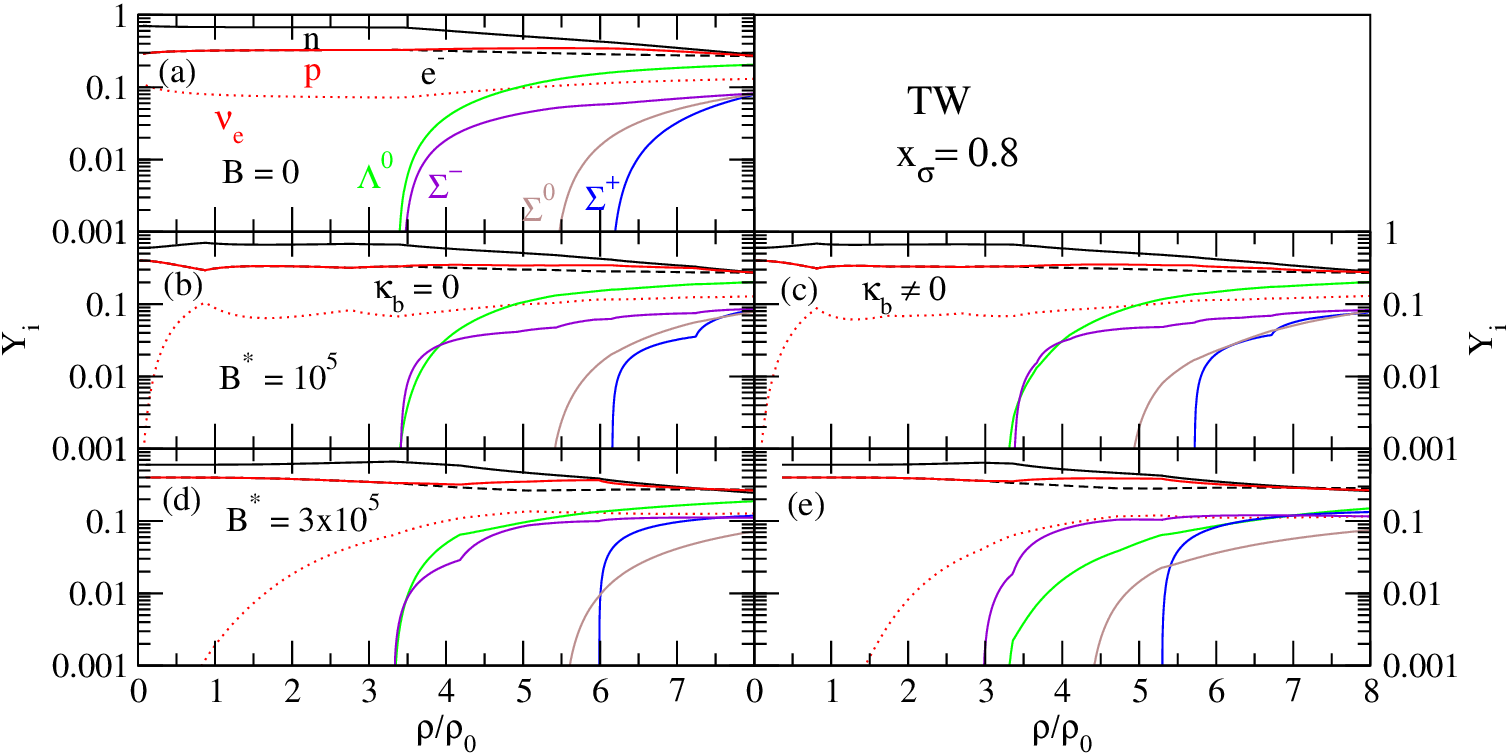}
\caption{Baryon fraction versus baryon density obtained with TW model, for $x_{\sigma} = 0.8$ and  neutrino-trapped matter.}
\label{frac08nt}
\end{figure}

\begin{figure}[htb]
\vspace{1.5cm}
\includegraphics[width=0.8\linewidth,angle=0]{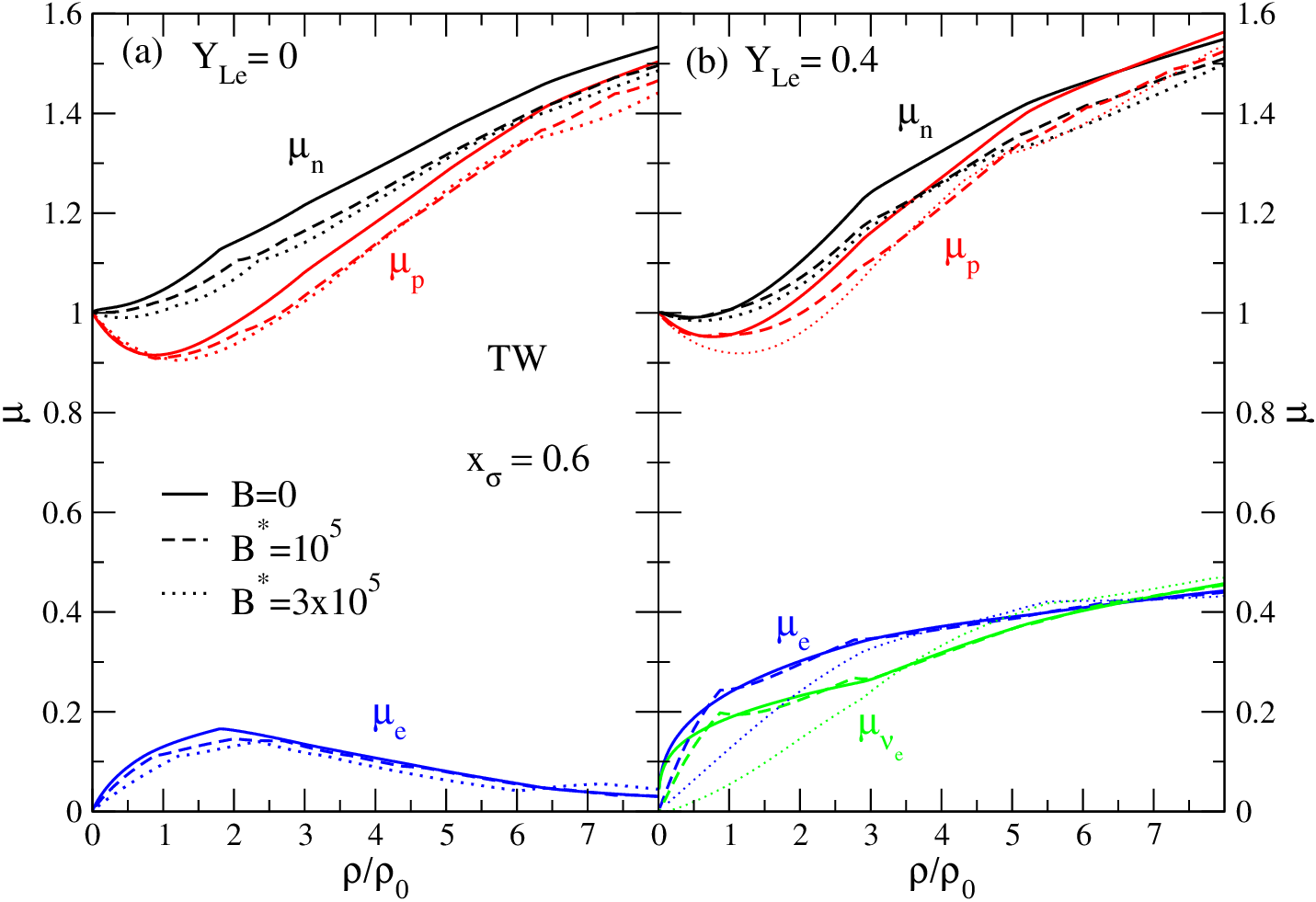}
\caption{Chemical potentials versus baryon density obtained with TW model, for
  several values of magnetic field, without  AMM, and for $x_{\sigma}=0.6$ and
  for (a) neutrino-free matter; (b) neutrino-trapped matter.}
\label{chemp}
\end{figure}

In Figs.~\ref{frac06nf}, ~\ref{frac06nt}, ~\ref{frac08nf}, and~\ref{frac08nt},
we present the particle fractions $\displaystyle Y_i={\rho^v_ i}/{\rho}$
as function of the baryon density $\rho/\rho_0$, obtained with TW model using
two values of hyperon-meson coupling constants and for several values of the
magnetic fields. Figs.~\ref{frac06nf} and~\ref{frac08nf} are for neutrino-free
matter and  Figs.~\ref{frac06nt} and~\ref{frac08nt} for neutrino-trapped matter.  
For GM1 we get similar results.

First, we study the results obtained with the  hyperon-meson
coupling constant equal to $x_{\sigma}=0.6$. For $B=0$ neutrino free matter,
shown in Fig.~\ref{frac06nf} (a), the proton fraction rises quickly 
with increasing density and reaches $\sim 0.1$ around 2$\rho_0$ before the
appearance of hyperons. $\Sigma^-$ is the first hyperon to appear. In
Fig.~\ref{frac06nf} (b) and (d), we show the results for $B^*=10^5$ and
$B^*=3\times 10^5$. At low densities, the fraction of nucleons and leptons
are significantly affected by the magnetic field. The Landau quantization
increases the proton abundance, and, therefore,  the electron abundance due to the
charge neutrality. 
In Fig.~\ref{chemp} we show for TW with $x_\sigma=0.6$ the neutron, proton,
electron and neutrino chemical potentials as a function of density for different values of
the magnetic field.  It is seen that the proton and neutron chemical potentials
decrease with the increase of B.  This is also the general tendency  of leptons
except for the larger densities. In the absence of AMM the reduction of the neutron
chemical potential with an increase of the magnetic field  is due to a
reduction of the isospin asymmetry. The AMM will further reduce the neutron
chemical potential. Landau quantization is the main explanation for a
reduction of the proton chemical potential.
 As a consequence the threshold for the
appearance of hyperons like $\Sigma^-$ and $\Lambda$ will occur at larger
densities \cite{prak02}. However, at densities above $\sim \,5 \rho_0$ the threshold
densities for $\Sigma^0,\,  \Sigma^+,\, \Xi^-$ occur at smaller densities
probably due to the smaller effective masses of these hyperons at large densities with strong magnetic fields.
 The inclusion of AMM,  Fig.~\ref{frac06nf} (c) and (e), produces an even larger hyperon
suppression at the lower densities.

\begin{table*}[H]
\caption{The threshold density  of hyperons in fm$^{-3}$ when AMM is not included.  for several values of magnetic field, for both GM1 and TW models and for neutrino free and neutrino-trapped matter. We also identify, between brackets,  the hyperon that appears at smaller densities. } 
\label{table4}
\begin{ruledtabular}
\begin{tabular}{ccllll}
$B^{*}$  & Models & \multicolumn{2}{c}{$x_{\sigma}=0.6 $} & \multicolumn{2}{c}{$x_{\sigma}=0.8$}  \\
\cline{3-4} \cline{5-6}
          &  & $Y_{Le}=0$ &  $Y_{Le}=0.4$&  $Y_{Le}=0$  & $Y_{Le}=0.4$   \\
\hline
  $0$              & TW   & 0.275 ($\Sigma^-$) & 0.441 ($\Lambda$) & 0.301 ($\Sigma^-$) & 0.514 ($\Lambda$) \\
                      & GM1 & 0.280 ($\Sigma^-$) & 0.419 ($\Lambda$) & 0.327 ($\Sigma^-$)  & 0.501 ($\Lambda$)\\
$10^{5}$     & TW   & 0.311 ($\Sigma^-$) & 0.438 ($\Lambda$) & 0.338 ($\Sigma^-$) & 0.516 ($\Lambda$, $\Sigma^-$) \\
                     &  GM1 & 0.301 ($\Sigma^-$) & 0.418 ($\Lambda$) & 0.340 ($\Sigma^-$)  & 0.503 ($\Lambda$)\\
$3\times10^{5}$    & TW & 0.358 ($\Sigma^-$) & 0.445 ($\Sigma^-$) & 0.385 ($\Sigma^-$) & 0.506 ($\Lambda$, $\Sigma^-$) \\
                                & GM1  & 0.373 ($\Sigma^-$) & 0.427 ($\Lambda$, $\Sigma^-$)  &  0.433 ($\Sigma^-$) & 0.496 ($\Lambda$) \\
\end{tabular}
\end{ruledtabular}
\end{table*}

\begin{table*}[H]
\caption{The threshold density  of hyperons in fm$^{-3}$ when AMM is  included.  for several values of magnetic field, for both GM1 and TW models and for neutrino free and neutrino-trapped matter. We identify, between brackets,  the hyperon that appears at smaller densities.} 
\label{table5}
\begin{ruledtabular}
\begin{tabular}{ccllll}
$B^{*}$  & Models &\multicolumn{2}{c}{$x_{\sigma}=0.6 $}& \multicolumn{2}{c}{$x_{\sigma}=0.8$}  \\
\cline{3-4} \cline{5-6}
          &  & $Y_{Le}=0$ & $Y_{Le}=0.4$&  $Y_{Le}=0$ & $Y_{Le}=0.4$   \\
\hline
  $0$              & TW    & 0.275 ($\Sigma^-$) & 0.441 ($\Lambda$) & 0.301 ($\Sigma^-$) & 0.514 ($\Lambda$) \\
                      & GM1  & 0.280 ($\Sigma^-$) & 0.419 ($\Lambda$) & 0.328 ($\Sigma^-$) & 0.501  ($\Lambda$) \\
$10^{5}$      & TW   & 0.291 ($\Sigma^-$) & 0.433  ($\Lambda$) & 0.315 ($\Sigma^-$) & 0.500  ($\Lambda$) \\
                                &  GM1  & 0.287 ($\Sigma^-$) & 0.409 ($\Lambda$, $\Sigma^-$) &  0.325 ($\Sigma^-$) & 0.488 ($\Lambda$) \\
$3\times10^{5}$    & TW & 0.344 ($\Sigma^-$) & 0.389 ($\Sigma^-$) & 0.373 ($\Sigma^-$) & 0.457 ($\Sigma^-$) \\
                                 & GM1  & 0.359 ($\Sigma^-$) & 0.375 ($\Sigma^-$) & 0.415 ($\Sigma^-$) &  0.468 ($\Sigma^-$)\\
\end{tabular}
\end{ruledtabular}
\end{table*}

We consider now matter with trapped  neutrinos. For
$B=0$,  Fig.~\ref{frac06nt} (a), the proton and electron fractions 
are $\sim 0.3$ from low densities until the appearance of hyperons at
$\sim 3\rho_0$ and, therefore, the neutrino fraction is $\sim 0.1$ since we
are imposing $Y_L=0.4$. $\Lambda$ is
the first hyperon to appear for B=0. For $B^*=10^5$ (see Fig.~\ref{frac06nt}
(b)), the proton fraction is larger at the smaller densities, $\sim 0.4$, 
decreases to 0.3 for $\rho\sim \rho_0$ and stays approximately constant  until the
appearance of hyperons at $\sim 3\rho_0$.  The neutrino fraction rises
quickly with increasing density and reaches $\sim 0.1$ around $\rho_0$ and
keeps, roughly, this value, $\sim 0.1$.  $\Lambda$ is still the first hyperon to appear in
this case. However, for $B^*=3\times 10^5$ (see Fig.~\ref{frac06nt} (d)), the
proton fraction decreases  slightly starting at  $\sim 0.4$ at lower  densities until the appearance of
hyperons at $\sim 3\rho_0$. $\Sigma^-$ is the first hyperon to appear almost at the same density as $\Lambda$. The neutrino fraction is almost zero at subsaturation densities, and
rises, quickly, with increasing density, to  $\sim 0.2$ around $6
\rho_0$. The net effect
of magnetic field is a neutrino suppression due to the larger proton  and therefore electron fractions and to  move the threshold density of the negatively charged baryons (\textit{e.g.} $\Sigma^-$) to larger densities and the positively charged baryons to lower densities. This is due to a decrease (increase) of the chemical potentials of negatively charged baryons (positively charged baryons) when the neutrino chemical potential is taken into account. In particular the onset of the $\Xi^-$ occurs for densities larger than 8$\rho_0$.  The neutral baryons are not affected by the presence of neutrinos. In the presence of a strong magnetic field the overall effect of the neutrinos is smaller than for B=0.

For $x_{\sigma}=0.8$, we obtain  similar results and the main differences are: at high densities  the onset of $\Xi^-$ occurs at $\rho<8\rho_0$, the onset of the $\Sigma^-$ and $\Lambda$ occurs at larger densities and hyperon fractions are smaller.

The effect of the magnetic field on the onset of hyperons, in both cases for neutrino free and neutrino trapped matter, is clearly shown in Tables~\ref{table4} and~\ref{table5}, where we give the threshold density of the first hyperon, for TW and GM1 models and for $x_{\sigma}=0.6, \:\hbox{and}\:0.8$, without including the AMM  (Table~\ref{table4})and including  the AMM in (Table~\ref{table5}).  The main conclusions for neutrino-free matter are: a)  the hyperon onset occurs with the appearance of the $\Sigma^-$ meson, both with and without AMM; b) the onset density generally increases with the increase of the magnetic field but for GM1 it may decrease for $B^*=10^5$;  c) the inclusion of AMM reduces the onset density of hyperons; d) the hyperon onset occurs at larger densities for $x_\sigma=0.8$.  For neutrino-trapped matter we conclude that: a)   the onset density of hyperons is generally get smaller when the magnetic field magnitude increases; b) for small fields $\Lambda$ is the first hyperon to appear but for large fields it is either $\Sigma^-$ or $\Sigma^-$ and $\Lambda$ together.

\begin{figure}[ht]
\vspace{1.5cm}
\centering
\includegraphics[width=0.85\linewidth,angle=0]{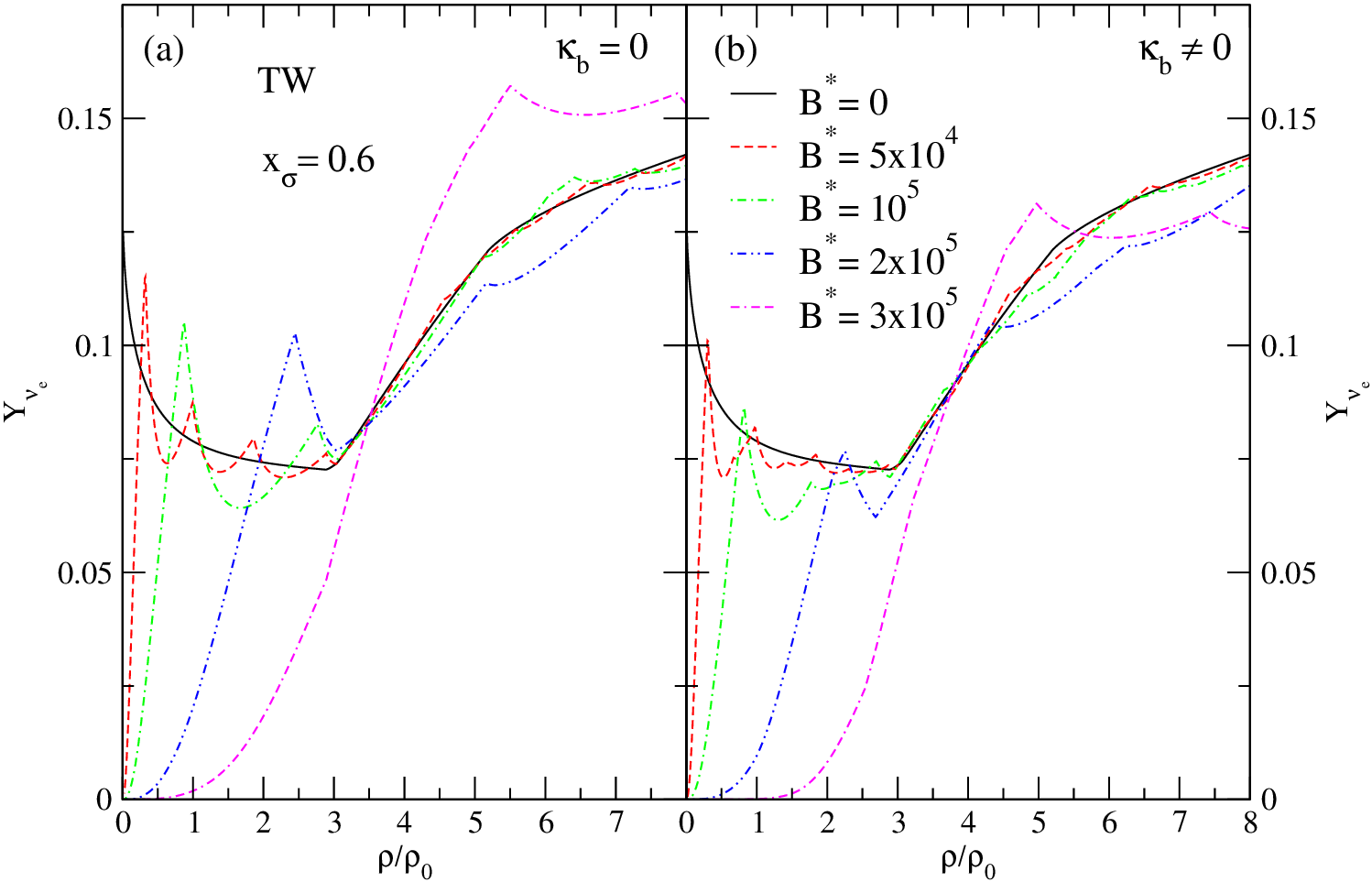}
\caption{Neutrino fraction versus baryon density obtained with TW model, for several values of the magnetic field without/with including AMM and for $x_\sigma=0.6$.}
\label{fracneut06}
\end{figure}

In order to better understand the effect of the magnetic field on the neutrino trapping, we show   in Fig.~\ref{fracneut06} the fraction of neutrinos for several values of magnetic field.  For $B=0$, the neutrino fraction decreases at low densities and starts to increase after the onset of hyperons because with the hyperon onset the electron fraction gets smaller. For $B^{*}=5\times 10^4$ and $B^{*}=10^5$, the neutrino fraction is affected by the Landau quantization of electrons, and oscillates around the $B=0$ results, when AMM are not included. At low densities the main effect of magnetic field is the suppression of neutrinos. For $B^{*} > 10^5$,  the neutrino fraction is lowered at low densities and enhanced at higher densities. However, when the AMM are included, the abundance of neutrinos is only slightly reduced for $B^{*}=5\times 10^4$ and $B^{*}=10^5$, while for $B^{*} > 10^5$ the neutrino suppression is strong, \textit{e.g.} for $B^{*}=3\times 10^5$ there are no neutrinos at a density below the saturation density. We conclude, therefore, that the abundance of neutrinos in the presence of  strong  magnetic fields has a  strong suppression at low densities.

In order to study the effect of the magnetic field on the properties of the stars described by the EOS discussed above with  neutrino trapping, we must consider a  magnetic field depending on the baryon density, which is not larger than $10^{15}$G at the surface. We will do this in the next section.

\subsection{Baryon density-dependent magnetic field}
Since, to date, there is no information available on the interior magnetic field of the star, 
we will assume that the magnetic field is baryon density-dependent as suggested by
Ref.~\cite{chakrabarty97}. The variation of the magnetic field $B$ with the
baryon density $\rho$ from the center to the surface of a star is
parametrized~\cite{chakrabarty97, mao03} by the following form
\beq
B\left(\frac{\rho}{\rho_0}\right)=B^{\hbox{surf}}+B_0\left[1-\exp\left\lbrace-\beta\left( \frac{\rho}{\rho_0}\right)^\gamma  \right\rbrace  \right],
\eeq 
where $\rho_0$ is the saturation density, $B^{\hbox{surf}}$ is the magnetic
field at the surface taken equal to $10^{15}$G, in accordance with the
values inferred from observations and $B_0$ represents the magnetic field at
large densities. The parameters $\beta $ and $\gamma$ may be chosen in such way that the field decreases fast or slowly with the density from
the center to the surface. In this work, we will use one set of value ($\beta=0.05$ and $\gamma=2$) allowing a slowly varying field. The magnetic field will be given 
in units of the critical field $B^c_e=4.414 \times 10^{13}$~G, 
so that $B_0=B^*_0 \, B^c_e$. We further take $B_0$
as a free parameter to check the effect of different fields. A detailed discussion 
about the variation of the magnetic field with the baryon density 
may be found in Ref.~\cite{aziz09}.

In this section, we will only  present the results for the density-dependent relativistic model TW  and we will take for the hyperon-meson coupling constants $x_{\sigma}=0.6$. Furthermore, in all the figures, we will only show the results obtained without including the baryonic AMM, because for the intensity of the magnetic fields considered its contribution to the EOS is negligible.

\begin{figure}[ht]
\vspace{1.5cm}
\centering
\includegraphics[width=0.85\linewidth,angle=0]{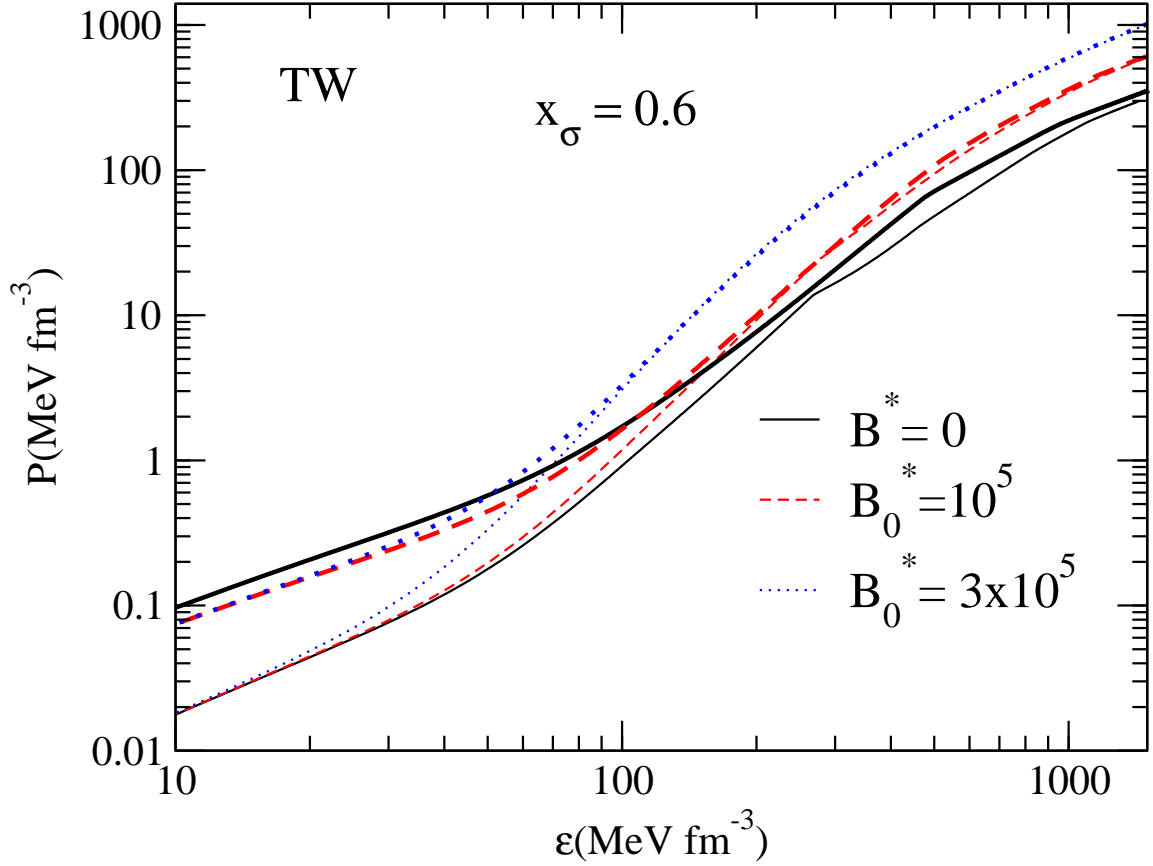}
\caption{EOS for stellar matter for TW model without/with neutrino trapping, for $x_\sigma=0.6$,   for several values of magnetic field and without AMM, using a slowly varying parametrisation of $B$. Thin lines correspond to neutrino-free matter and thick lines to neutrino-trapped matter.}
\label{eostwbv}
\end{figure}
In Fig.~\ref{eostwbv}, we show the total pressure $P$, see Eq. (\ref{press1}), as a function of the
total energy density $\varepsilon$ for the magnetic field strengths $B^*=0, 10^5$, and $3 \times 10^5$. The results with neutrino-trapped and
neutrino-free matter are plotted with thick and thin lines, respectively. The
EOS with neutrino trapping is stiffer than the neutrino free EOS, both with
and without magnetic field. However, when neutrino trapping is imposed  the
magnetic field makes the EOS softer at low density. 

\begin{figure}[ht]
\vspace{1.5cm}
\centering
\includegraphics[width=0.85\linewidth,angle=0]{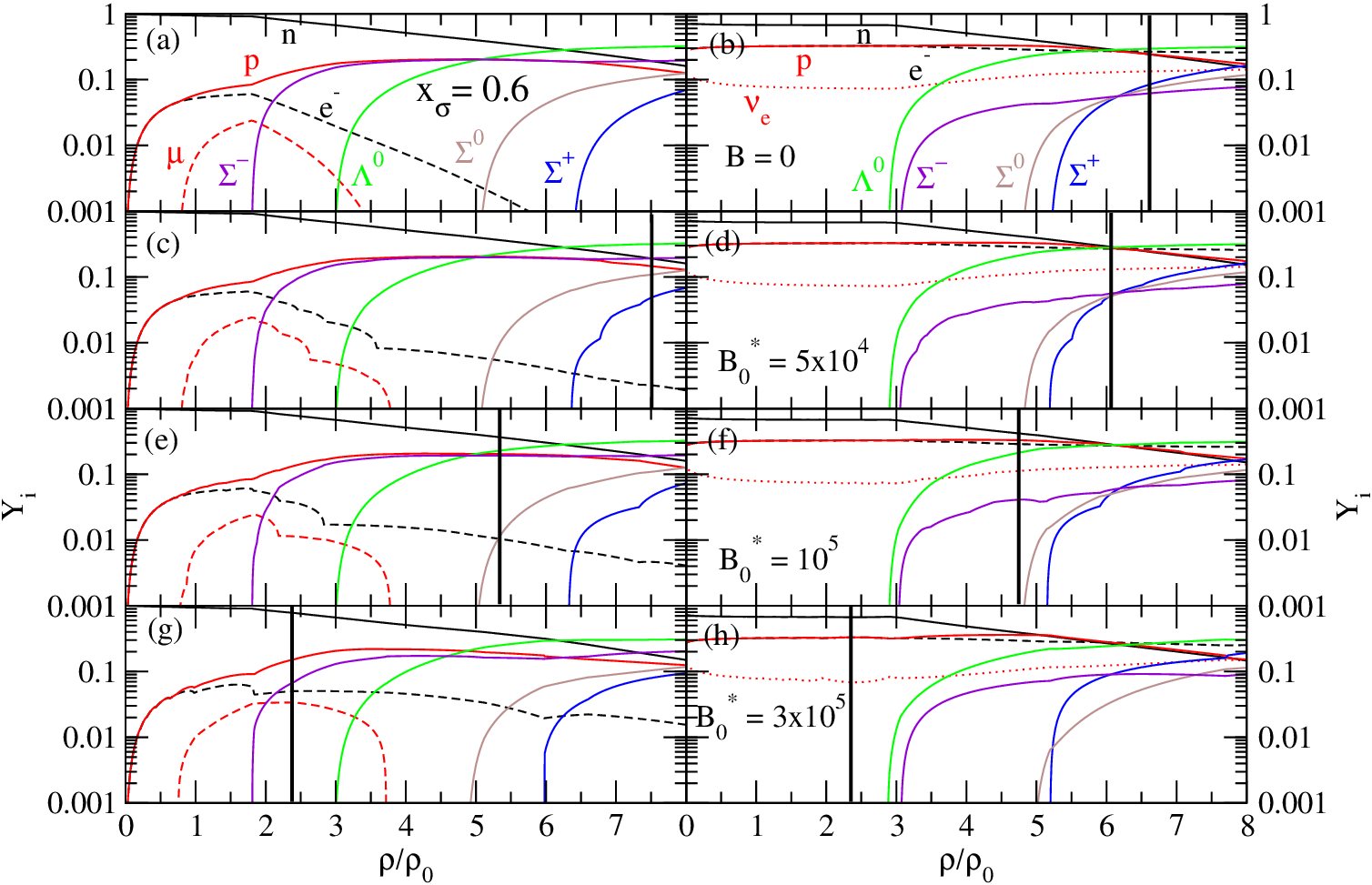}
\caption{Baryon fraction versus density obtained with TW model with/without
neutrino trapping, for several values of the magnetic field, using a slowly
varying parametrisation of $B$, and without AMM. Figures (a), (c), (e), and
(g) are for neutrino free and (b), (d), (f), and (h) for neutrinos
trapped.}
\label{fracbv}
\end{figure}

We have also studied the baryonic and leptonic composition of the stars.
In Fig.~\ref{fracbv} we plot the lepton and the  baryon fractions for neutrino
free matter (left column) and for neutrino trapped matter (right column) for TW model. 
We show the particle 
fractions for the smallest and the largest magnetic fields considered in the present work. 
The vertical lines in the figures represent the central density of the star configuration 
with maximum mass when it lies within the range of densities shown. 
We conclude that the main effect of  magnetic fields with the intensity
considered is the appearance 
at high densities of a larger leptonic fraction.  Due to the contribution of
the magnetic field energy and pressure to the total EOS of stellar matter the
contribution of matter becomes smaller when the magnetic field increases and,
therefore, the fraction of strangeness in the star is strongly reduced because the larger contributions of hyperons for the star come from the larger densities.


\begin{figure}[ht]
\vspace{1.5cm}
\centering
\includegraphics[width=0.85\linewidth,angle=0]{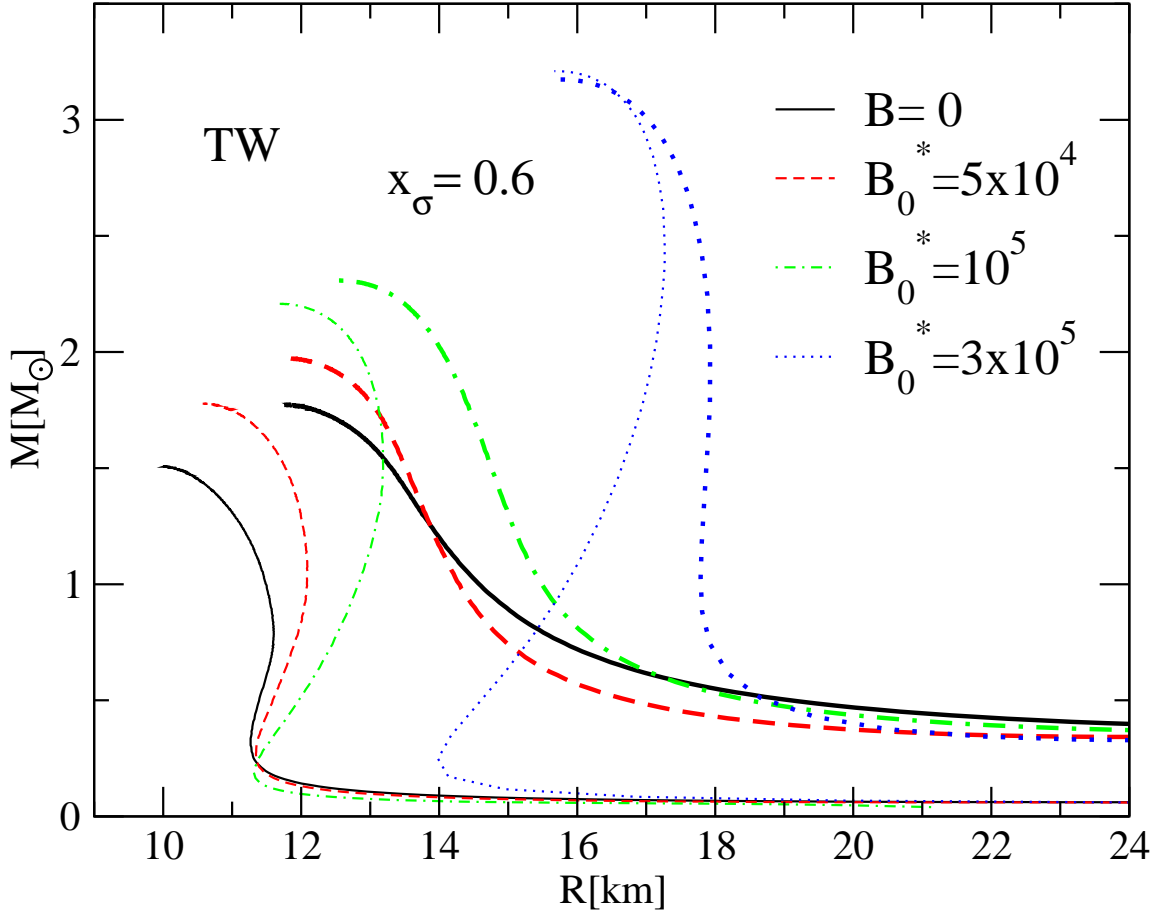}
\caption{Mass-radius curve of neutron stars for several values of the magnetic
  field, using a slowly varying parametrisation of $B$, for TW model
  with/without neutrino trapping, and for $x_\sigma=0.6$. Thin lines
  correspond to  neutrino free matter and thick lines to  neutrino trapped matter.}
\label{massgrav}
\end{figure}
Hadron star profiles are  obtained from the EOS studied, for severals values
of magnetic field, by solving the Tolman-Oppenheimer-Volkoff equations,
resulting from Einstein's general relativity equations for spherically
symmetric and static stars. Although an approximation due to the presence of the magnetic field, we assume  spherical
symmetric stars. In Table~\ref{table6} we show the values obtained for the
maximum gravitational and baryonic masses of the stars, their radius, their
central energy and  baryon densities and the  magnetic field at the center.
 The results are shown for the stars with and without trapped neutrinos. The
 mass/radius curves for the families of stars corresponding to the maximum
 mass configurations given in Table~\ref{table6} are plotted in
 Fig.~\ref{massgrav}; the thick lines correspond to the stars with  the
 trapped neutrinos. 
From Table~\ref{table6}, we conclude that maximum mass and radius of a neutrino free star is
generally smaller than the corresponding mass and radius of a star with
trapped neutrinos.  The presence of neutrinos makes the EOS stiffer.
This difference, however, is reduced with the increase of
the magnetic field. Quantitatively, in the field free case the difference is around
$1.75$ km and it decreases to $1.22$ km for $B^*_0=5\times 10^4$, to $0.85$ km
for $B^*_0=10^5$, and to $0.09$km $B^*_0=3\times 10^5$. This is due to the
fact that the contribution of matter becomes smaller and smaller as the
magnetic field increases, and therefore the mass and radius are not so
sensitive to the specific properties of the stellar matter contribution to the
total EOS. It has already been shown in several works~\cite{prak02, mbap09, aziz09}
that the stronger the magnetic field the stiffer the EOS and therefore the larger the star radius and mass.

An analysis of the Table~\ref{table6} allows us to draw some interesting
conclusions: a) although the presence of neutrinos makes the maximum
gravitational and baryon masses larger, the difference with respect to the   maximum
gravitational and baryon masses of a neutrino free star decreases with the
increase of the magnetic field intensity.  For the largest field considered
the star with neutrinos has  a smaller mass; b) as a consequence in the
presence of a very strong magnetic field a star will never decay into a
blackhole after the outflow of neutrinos. This, however, may occur if no
magnetic field/a weak magnetic field exists because the baryonic mass of the
maximum mass star configuration is larger if neutrinos are trapped in the
star; c) the most massive neutron star reported till recently was the
PSR J1748-2021B with a mass 2.74$\pm 0.21$~\cite{freire08}, which, however, was
still not confirmed by other measurements.  This kind of measurements
may impose a constraint on the maximum acceptable magnetic field. For
$B^*_0=3\times 10^5$ we get at the center of the star $B\sim 4\times 10^{18}$G
and a mass larger than 3.2 M$_\odot$.

\begin{table*}[H]
\caption{Compact star properties using TW model,  for several values of a density-dependent
magnetic field. $M_{max}$, $M^b_{max}$, R, $E_{0}$, $\rho^c$, and $B^*_c$
are the gravitational and baryonic masses, the radius, the  central energy
and baryon density, and the value of the magnetic field at the center, respectively.} 
\label{table6}
\begin{ruledtabular}
\begin{tabular}{lccccccc}
&$B^{*}_{0}$ & $M_{max} [M_\odot]$ & $M^b_{max} [M_\odot]$ & R [km] & $E_{0}[\hbox{fm}^{-4}]$ & $u^ c=\rho^c/\rho_0$ & $B^*_c $     \\
\hline
$Y_{Le}=0.4$   &   $B^{*}=0$        & 1.77 & 1.95 & 11.75 & 6.35 & 6.627 &  0   \\
$Y_{\nu_e}=0$ &   $B^{*}=0$        & 1.51 & 1.70 & 10.00 & 7.74 & 8.460 &   0  \\
$Y_{Le}=0.4$   &$5\times 10^{4}$ & 1.97 & 2.17 & 11.81 & 6.10 & 6.041 & 4.196$\times10^{4}$ \\
$Y_{\nu_e}=0$ &$5\times 10^{4}$ & 1.78 & 2.00 & 10.59 & 7.25 & 7.514 & 4.705$\times10^{4}$ \\
$Y_{Le}=0.4$   & $10^{5}$             & 2.31 & 2.51 & 12.54 & 5.34 & 4.740 & 6.750$\times10^{4}$ \\
$Y_{\nu_e}=0$ & $10^{5}$             & 2.21 & 2.45 & 11.69 & 5.82 & 5.303 & 7.551$\times10^{4}$ \\
$Y_{Le}=0.4$ & $3\times 10^{5}$  & 3.18 & 3.28 & 15.76 & 3.22 & 2.365 & 7.323$\times10^{4}$ \\
$Y_{\nu_e}=0$ &$3\times 10^{5}$  & 3.21 & 3.43 & 15.67 & 3.15 & 2.376 & 7.379$\times10^{4}$ \\
\end{tabular}
\end{ruledtabular}
\end{table*}

\section{Conclusions and outlooks}
In the present work we have studied the effect of a strong magnetic field on
the EOS and the star properties when neutrinos are trapped in the star. We
have used two different relativistic mean-field  models, one with constant
coupling parameters, GM1, and the other with density dependent coupling
parameters, TW. We have also considered two sets of hyperon-coupling
parameters. The main conclusions of the work  do not depend either on the
model or on the strength of the hyperon-couplings.

The phase of trapped neutrinos in the life of a proto-neutron star
occurs while the stellar matter is still warm \cite{burrows86,prak97}, and,
therefore, a finite temperature calculation should have been done. However, we do not
expect that temperature will change the main conclusions of the present
work. In fact, in several works \cite{prak97,mp03,mp04} it has been shown that
the star properties such as mass and radius do not depend much on temperature.

We have shown that a strong magnetic field suppresses the presence of
neutrinos at low densities. It was also shown that although strangeness 
is suppressed by the presence of neutrinos, if the star has a strong magnetic 
field this suppression is smaller. The magnetic field affects in a different way the
charged and neutral baryons and it may affect the order at which density they
appear. For neutrino free matter, $\Sigma^-$ is the first hyperon to appear 
at the smallest densities. However, for neutrino trapped matter, $\Lambda$ 
is the first hyperon to appear except for the largest field considered when 
we may have $\Sigma^-$  or $\Sigma^-$ and $\Lambda$.

We have studied the properties of stars with trapped neutrinos and strong
magnetic fields: it was shown that the magnetic field increases the mass and
radius of the most massive star configuration, {and in general the radius of
all stars with a mass larger than $0.5$ M$_\odot$.} The mass of observed  neutron
stars may set an upper limit on the possible magnetic field acceptable in the
interior of a star. The contribution of the
magnetic field to the total EOS of the star reduces the relative importance
of the stellar matter term: the central baryonic density decreases
as the field increases. One of the main consequences is the reduction of the
strangeness fraction in the star or other exotic components such as a quark core \cite{aziz09}, or kaon condensation \cite{shen08}.  It was also shown that the magnetic field
reduces the possibility of formation of a blackhole after the outflow of neutrinos.

\begin{acknowledgments}
We would like to thank Joao da Provid\^encia for many helpful and
elucidating discussions. This work was partially supported by FEDER and Projects PTDC/FP/64707/2006 and CERN/FP/83505/2008, and  by COMPSTAR, 
an ESF Research Networking Programme.
\end{acknowledgments}

\end{document}